\documentclass{article}

\usepackage{arxiv}

\usepackage[utf8]{inputenc} 
\usepackage[T1]{fontenc}    
\usepackage{hyperref}       
\usepackage{url}            
\usepackage{booktabs}       
\usepackage{amsfonts}       
\usepackage{amsmath}
\usepackage{nicefrac}       
\usepackage{microtype}      
\usepackage{chemfig}		
\usepackage{graphicx}
\usepackage[style=numeric-comp,sorting=none]{biblatex}
\bibliography{references}  
\usepackage{doi}
\usepackage{authblk}
\usepackage{subcaption}
\usepackage{xcolor}

\title{Transferability of datasets between Machine-Learning Interaction Potentials}

\date{} 					

\author[1]{Samuel P. Niblett}
\author[2]{Panagiotis Kourtis}
\author[2]{Ioan-Bogdan Magd\u{a}u}
\author[1]{Clare P. Grey}
\author[3]{G\'abor Cs\'anyi$^*$}

\affil[1]{Yusuf Hamied Department of Chemistry, University of Cambridge, Lensfield Road, Cambridge, UK}
\affil[2]{School of Natural and Environmental Science, Newcastle University, Newcastle upon Tyne, NE1 7RU, UK}
\affil[3]{Engineering Laboratory, University of Cambridge, Trumpington St and JJ Thomson Ave, Cambridge, UK}

\begin{document}
\maketitle

\begin{abstract}
With the emergence of Foundational Machine Learning Interatomic Potential (FMLIP) models trained on extensive datasets, the question of how far data can be transferred between different ML architectures has become increasingly important. In this work, we examine the extent to which training data optimised for one machine-learning forcefield algorithm may be re-used to train different models, aiming to accelerate FMLIP fine-tuning and to reduce the need for costly iterative training.
As a test case, we train models of an organic liquid mixture that is commonly used as a solvent in rechargeable battery electrolytes and that plays an important role in degradation processes of these devices, making it an important and representative target for reactive MLIP development.
We assess the performance of our models by analysing the stability and thermodynamic accuracy of molecular dynamics trajectories, showing that this is a more stringent test than comparing prediction errors for particular configurations.

We consider several types of training configuration, and several popular machine-learning protocols - notably the recent MACE architecture, a message-passing neural network designed for high efficiency and smoothness. We demonstrate that simple training sets constructed without any {\it ab initio} dynamics simulations are sufficient to produce stable models of molecular liquids that can transfer to multiple liquid compositions. For simple neural-network architectures, further iterative training is required to capture the thermodynamic and kinetic properties of the liquid correctly, but MACE appears to perform well with extremely limited datsets. We find that configurations which are designed by human intuition to correct systematic deficiencies of a model are effectively transferred between algorithms, but that active-learned data that are generated by one MLIP do not typically benefit a different algorithm. As in other tests, MACE shows better performance with transferred active-learned data than traditional neural networks do.
Finally, we examine the effect of transferred dataset size on a model's ability to generalise to unseen molecules. We find that any training data which improve model performance for the base molecule also improve stability for related unseen molecules, suggesting that trajectory failure modes are connected with chemical structure rather than being entirely system-specific.

These results provide insight into how training set properties affect the behaviour of an MLIP, and practical principles to assist rapid enhancement of training sets for forcefields of molecular liquids. These approaches may be used in tandem with foundation models to dramatically accelerate the rate at which new chemical systems can be studied by these methods.
\end{abstract}

\section{Introduction}

Recent years have seen an explosion in the field of atomistic simulation for materials and molecular liquids, 
driven both by the burgeoning importance of electrochemical\cite{CanoChen2018,GreyHall20,SIB_roadmap_21,Supercaps,electrocatalysts} and nanostructured devices\cite{BocquetNanofluidics,MOFs} to enhance sustainable technology and by the emergence of machine-learning interaction potentials (MLIPs) as a simulation method that approaches quantitative accuracy for comparison with experiment. MLIPs represent an effective compromise between the efficiency of classical molecular dynamics and the accuracy of {\it ab initio} quantum calculations, 

facilitating simulations of complex molecular systems and materials at an electronic-structure level of accuracy for the first time.\cite{musil2021physics,deringer2021gaussian,langer2022representations} A particularly exciting achievement is the emergence of models that can describe bond-breaking chemical reactions on time- and length-scales needed to simulate complex environments such as electrolytes and solid/liquid interfaces.\cite{behler2017first,GalibLimmer2021,Stark2024,YangRoitberg2024} 

Early MLIP models were found to give excellent agreement with reference quantum calculations for configurations near their training set, but were often poor at extrapolating their energy and force predictions to describe unseen atomic environments, and therefore unstable in MD simulations.\cite{behler2017first} This deficiency, combined with the high cost of obtaining data and training models, means that most MLIPs to date have been limited to a small region of chemical space. Their training datasets attempt to mimic an ensemble probability distribution for a particular target system or systems, so that the model is rarely required to extrapolate to unseen configurations. Training set development then becomes a sampling problem where the potential energy surface being sampled is too expensive to evaluate for a significant number of configurations.

Recently, the foundational model method has disrupted this approach: the advent of efficient and smooth ML algorithms (particularly those based on equivariant graph neural networks)\cite{MPNNs,MACE} has made it possible to train on large and diverse datasets to produce models that can capture a large and diverse chemical space.\cite{CHGNet, SevenNet,M3GNet,MACE_MP_0,MatterSim} These foundational models can provide robust molecular dynamics simulations for many different applications,\cite{MACE_MP_0}, but they often fall short of quantitative accuracy in computing thermodynamic or kinetic properties. 
To use them in a predictive capacity, additional training data specific for the target system must be added - this helps to specialise (or fine-tune) the model to solve the desired problem.\cite{MACE_MP_0,FocassioSchleder2024,Deng24} Understanding the extent to which different training sets may profitably be combined and mixed is a crucial component of developing a reliable fine-tuning strategy.

Training set optimisation, whether to fine-tune a foundational model or to develop a standalone MLIP for a particular system, requires selecting additional training data (e.g. by intuition or by observation) to cover regions of configuration space where the initial model performs poorly. This process is typically performed in stages with particular model deficiencies targeted at each stage, so we refer to it as ``iterative training''. An important special case of iterative training is ``active learning'', where the MLIP itself is used to generate training configurations through MD or Monte Carlo approaches. Typically one selects the MD configurations with the highest prediction errors, usually estimated using a committee error analysis,\cite{behler2017first,SchranMarsalek2020} for incorporation to the training set.

Active learning procedures can be very effective, but also tedious and expensive. Many configurations sampled from MLIP trajectories will provide little or no benefit to the  training set and hence represent wasted computation - especially since there is no guarantee that the configuration space of an initial MLIP is a good approximation to the target space. Minimising the amount of iterative training that must be performed is therefore highly desirable, and data re-use is an appealing route to achieving this reduction. As the MLIP field expands, researchers will increasingly find that the system they are interested in has already been studied, but often with a different ML algorithm or set of hyperparameters than the one they intend to use. Thanks to many funders' Open Data policies, the datasets for these models are often publicly available. Therefore simply training a new model on the existing dataset might offer a quick route to a usable MLIP (or at least reduce the number of iterative training generations required to obtain one). However, data that were tailored to improve the performance of one MLIP algorithm may not be useful or relevant to another. This paper aims to explore the opportunities and limitations of re-using data in this way.

We base our investigation around three questions: 
\begin{enumerate}
    \item How should performance of MLIPs be quantified/compared, and what magnitude errors are tolerable in initial development of a model?
    \item What types of training configuration provide most benefit to MLIP performance, and can these configurations be transferred between different algorithms? 
    \item Which types of training configuration provide most benefit to the generalisation ability of an MLIP (i.e. its ability to describe chemical systems not included in the training set)? 
\end{enumerate}

We explore each of these questions by training MLIPs for an organic liquid mixture: ethylene carbonate (EC) and ethyl methyl carbonate (EMC) (structures in \ref{fig:SIstructures}). Mixtures of these molecules are commonly used as solvents for lithium-ion battery electrolytes.\cite{Xu14} Reactive decomposition of these solvents is a key step in degradation mechanisms that limit battery lifetime,\cite{RinkelGrey20} so simulating these two molecules (and ultimately their reactions) is an important task that has already received significant attention from the MLIP community.\cite{dajnowicz2022high,magduau2023machine,BAMBOO} %

Moreover, the mixture has several properties that make it particularly interesting and challenging to describe computationally: large molecular dipoles, a combination of stiff and soft intramolecular degrees of freedom, and the ability to sample a range of local compositions (ranging from pure EC to pure EMC in a given volume of liquid). Developing MLIPs for liquids such as these is therefore a challenging but valuable computational effort, and understanding data transferability for these systems is a rewarding objective.

Battery development research often involves comparing properties of multiple cells with distinct but related electrolytes. To emulate this workflow it would be desirable to have a single MLIP capable of describing a range of electrolyte molecules, without needing expensive explicit training for each. It is therefore instructive to consider how our models generalise to out-of-distribution molecules, and whether some types of training configuration confer more benefit for this task. In sec.~\ref{sec:Other_molecules}, we consider how MLIP models trained to describe EC:EMC mixtures perform when tasked with simulating EC:DEC, VC:EMC, and PC - further common electrolyte molecules.

Generalisability of models to study multiple chemical systems has been studied previously\cite{KandyLam23,GoodwinEtAl24,BAMBOO,GouldVuckovic24} and various model types have been shown to be particularly effective. However, to our knowledge this paper is the first to investigate the efficacy of transferring data to different types of models and molecules simultaneously, and to consider how generalisability depends on training set as well as model architecture.

Sec.~\ref{sec:background} further discusses the importance of training sets in MLIP design, and provides a brief overview of how they are generated. Sec.~\ref{sec:methods} presents our approach to investigating data transferability and assessing MLIP performance for molecular liquids. In sec.~\ref{sec:gen0results} we compare how two different types of MLIP perform on transferred datasets, finding that message-passing graph neural networks give good descriptions of molecular liquids even for very small training sets and are better able to utilise transferred data than traditional neural network models. We explore the effect of transferred data on active-learning efficiency in secs.~\ref{sec:ALresults} and \ref{sec:VolScans}, and on chemical generalisation in sec.~\ref{sec:Other_molecules}.

\section{Background}
\label{sec:background}

\subsection{Training set design in MLIPs}

The poor extrapolative ability of most MLIP algorithms poses a particular and serious challenge to MD simulations of liquids. The high dimensionality of the configuration spaces for these systems makes them extremely difficult to sample and the energy often varies considerably over small parameter ranges. In principle, a huge training set would be able to capture all the important features of a liquid configuration space, and hence allow us to run MD in the interpolative regime only. But generating such training data would be difficult and exceptionally expensive, and training a model on such a large training set would be a demanding task in its own right. Instead, the objective is usually to develop a minimal training set for an MLIP that reproduces known reference data correctly (for example, densities and diffusivities of a liquid) and then to compute related quantities that are inaccessible to reference quantum calculations (for example, solution conductivity). 

However, one cannot control or know in advance which atomic environments will be explored by a dynamic simulation, so it is impossible to ensure that the simulated trajectory does not leave the interpolative region of the model. Even in the impractically expensive case where the training set includes all high-probability regions of the target configuration space, an MLIP MD simulation will eventually leave its interpolative region if it under-predicts the energy of an unphysical high-energy configuration (either systematically or stochastically). 
Once this excursion has happened, there is no guarantee that the MLIP has correctly learned the restoring force to drive the system back into the well-trained region, often leading to catastrophic failure of the trajectory. Even without such a failure, regions of the potential energy surface with high errors can exert a dramatic effect on the configurational probability distribution for the model.\cite{BehlerCsanyi2021} 

The usual procedure to mitigate these issues is to use iterative training (particularly active learning) to fill in the ``holes'' left in an initial training set - particularly by training on a subset of the high-energy configurations to prevent their energies being under-predicted.\cite{behler2017first} Therefore a key objective of the active learning process is to teach a model which regions of configuration space are inaccessible under simulation conditions, i.e. training on unfavourable high-energy configurations to place ``barriers'' that restrict a model from leaving the high-probability regions of its phase space.

These low-probability configurations are usually poorly represented in initial training sets - which by construction aim to reproduce the equilibrium distribution of state occupancies, not the low-probability tails of this distribution. This shortcoming is particularly true of training data obtained from a long AIMD simulation, often considered the gold standard for training MLIPs of liquids.\cite{GalibLimmer2021}
 Configurations sampled from a classical MD trajectory (or other auxiliary model) will usually not represent even the high-probability configurations of the reference method. We typically expect that such datasets will not include sufficient information to prevent a model from accessing unphysical space, thus resulting in highly unstable and inaccurate simulations.

We expect that MLIPs with different functional forms (i.e.~different algorithms) that are trained on the same dataset will have different-sized errors in each under-trained region of configuration space and thus will ``fail'' for different configurations and in different ways. Therefore, the iterative-training configurations required to correct these failures may be different for different MLIP algorithms. We hypothesise that a dataset optimised by iterative training for one class of model may not contain sufficient information to prevent another model from exploring unphysical configurations - i.e.~ active-learned configuration should not transfer well between models.

We aim to test this expectation by examining the extent to which training configurations known to be beneficial for one MLIP improve model accuracy for a different type of MLIP. Our analysis will include both active-learned configurations, the most common approach to refining an initial training set, and manually-generated configurations that are designed to correct specific deficiencies of a model. In order to consider the nature of these configurations, it is first necessary to discuss the different types of MLIP examined in our work.

\subsection{Classes of MLIP}

A great many MLIP protocols have emerged in recent years, based on almost every type of machine-learning algorithm.\cite{BehlerCsanyi2021} 
Typically, they attempt to learn a mapping between a feature space that encodes for different atomistic configurations, and the corresponding potential energies and force vectors. Those energies and forces then provide input for classical simulation methods.

The performance of MLIP protocols varies considerably but is very difficult to assess generally, due to the vast number of methods to compare and because different types of model have different advantages and problems to which they are best suited. However, this diversity of approaches and of model qualities (real or perceived) makes the ability to recycle data from one protocol to another even more important: even where a well-trained model for a particular problem exists, other groups may wish to use the same data with a different protocol for reasons of efficiency, expertise, or hardware compatibility. 

We consider data transferability between three MLIP methods representing quite different classes of learning algorithm. These are:
\begin{enumerate}
    \item GAP\cite{bartok2010gaussian}, a Gaussian Process regression model that represents atomic configurations using the turbo SOAP (Smooth Overlap of Atomic Positions) descriptors.\cite{caro2019optimizing}
    \item DeePMD\cite{DeePMD}, an end-to-end feed-forward neural network where both encoding to the feature space and the regression to atomic forces are performed by multilayer perceptrons.
    \item MACE,\cite{MACE} an equivariant message-passing neural network that combines many-body spatial features with a graph neural network regression.
\end{enumerate}

The precise features of these models will be discussed where they become relevant in the text. 
The main point to note here is that these models use three different types of symmetrised representation (turbo SOAP, neural-network embeddings of radial and angular information, and many-body atomic clusters) and three very different regression algorithms (gaussian process learning, neural networks, and messaging passing neural networks). 
Our results are therefore not restricted to a particular class of models, and our conclusions appear to be quite generally applicable.

\section{Methodology}
\label{sec:methods}

\subsection{Transferability of different training configuration types}
\label{sec:config_types}

The core idea of this work is an ablation study to compare the performance of MLIPs trained on different subsets of data from a complex optimised training set, to understand which types of configuration are the most transferable and what kinds of information they provide to a new model. 
The training set that we use for this purpose is that reported in ref.~\cite{magduau2023machine},
  which was developed to train a GAP MLIP for the EC:EMC mixture at various compositions. This dataset avoids the use of {\it ab initio} MD simulations, significantly reducing the initial cost to produce it, and contains around 1000 configurations - quite a small training set by the standard of molecular MLIPs. 

The GAP training set calculated reference energies using the plane-wave implementation of Castep\cite{stewart2005castep} using the PBE exchange-correlation functional with D2 dispersion correction (G06 keyword),\cite{grimme2006semiempirical} a plane-wave energy cutoff of 800 eV, and a Monkhorst-Pack 1 x 1 x 1 k-point grid. Standard CASTEP ultrasoft pseudopotentials were used to model the core electrons. The training set contained the following types of molecular configuration:
\begin{enumerate}
    \item 200 liquid configurations obtained from molecular dynamics simulations using the OPLS-AA forcefield\cite{OPLS_AA} at a constant composition of 4 EC molecules and 8 EMC molecules. 120 configurations were taken from NPT simulations at various elevated temperatures and pressures, and 80 from NVT simulations at 300-400\,K. The NVT set covers a range of state points, including densities as low as half the equilibrium value.
    \item Isolated molecules of EC and EMC in a variety of near-equilibrium configurations (200 configurations of EC and 400 of EMC)
    \item Volume scan data: 90 configurations obtained by taking frames from an initial GAP-MD trajectory and isotropically expanding or contracting them to obtain varying density, as discussed in ref.~\cite{magduau2023machine}. Note that these volume scans treat individual molecules as rigid (i.e. their bond lengths are unchanged), it is only the intermolecular separation that is varied by deformation of the box. The original dataset contained configurations with multiple compositions, but to focus on the effect of the volume-scan method we included only 20 configurations (3 distinct volume scans) that have the target composition of 4 EC and 8 EMC molecules. 
    \item 150 configurations obtained by 12 generations of iterative training using the GAP potential. At each generation 5-20 configurations with 
    anomalous densities or diffusivities were selected and added to the training set. 
    These configurations span a range of densities and compositions, but all contain a total of 12 molecules.
\end{enumerate}

To assess the transferability of each type of configuration, we train a series of MLIPs using DeePMD and MACE whose initial training data is taken from different subsets of the GAP dataset. Specifically we trained models using only dataset 1 ("OPLS only"), sets 1 and 2 ("OPLS+SM"), sets 1 and 3 ("OPLS+VS"), and all four sets combined ("Full"). Note that only the "Full" set samples all possible EC:EMC compositions, and only this set contains configurations that were active-learned by the GAP model.

We performed two pieces of analysis on our transferred models. The first was to assess the performance of MLIPs trained naively on transferred data alone, using the metric introduced in the following section. 
The second was to perform active learning on each transferred DeePMD model, to determine how many algorithm-specific configurations are required to correct the performance back to the GAP level in each case. Our hypothesis was that models that were initialised with more useful recycled data would have fewer unseen regions in the equilibrium configuration space, and hence require fewer generations of active learning to achieve full coverage of this space.

\subsection{Assessing model quality for liquid-state properties}
\label{sec:metric}

MLIPs are often benchmarked on fixed testing sets by comparing their prediction errors for energies and forces relative to the reference {\it ab initio} method. These test configurations are either held back from the original training set or sampled by MD carried out with the model being tested.
This approach essentially quantifies performance within a restricted volume of phase space and thus provides a controlled comparison of quality for different MLIPs. However, errors computed this way do not necessarily correlate with the ability of a model to reproduce the reference system's true properties. A more relevant, but more expensive, test of model quality is therefore to look at the predicted properties of the system directly. For a liquid, we consider three properties: trajectory stability, liquid density, and self-diffusivity. For simplicity, we compare properties for a single composition (3:7 EC:EMC) with a fixed cell size of 640 atoms.

Computing these equilibrium properties requires long (at least ns) molecular dynamics to achieve an adequate statistical sample, so the ability to propagate such trajectories without encountering significant unphysical behaviour is the most fundamental property required of an MLIP. We diagnose this ``trajectory stability'' by measuring the time period, $t_{\rm stab}$, that may be simulated without encountering dramatic changes in molecular structure (i.e. alterations in molecular connectivity or geometry) and liquid structure (e.g. evaporation) that would not occur in equivalent simulations using the reference DFT method. 
We define $t_{\rm stab,NPT}$ as the simulated time elapsed before an isothermal-isobaric trajectory triggers one of two failure conditions: the density leaves the range [0.2,2.0]g/cm$^3$, or one or more of the initial chemical bonds exceeds its equilibrium value by 0.5\,\AA. 
An equivalent quantity, $t_{\rm stab,NVT}$, describes the stability of canonical trajectories - in this case the density criterion is ignored.

Meeting either of these failure conditions indicates a catastrophic failure of the model, and we have observed that trajectories rarely if ever recover from such a position. In order to test stability very strictly, we perform our validation trajectories at 500\,K, a temperature at which high-energy configurations are comparatively easy to access. A MLIP that is stable at this temperature must have achieved small prediction errors for a large volume of the liquid phase space, in order to correctly disfavour unphysical configurations.

Liquid density and self-diffusivity are natural measures of the thermodynamic and dynamic properties of a liquid. They are simple to compute from a sufficiently long MD trajectory, strongly correlated with more complex liquid properties including conductivity, and sensitive to both low- and high-energy configurations of the liquid. The ability of an MLIP to reproduce these properties is therefore an appropriate measure of its accuracy. 

As mentioned previously, our MLIPs are trained against DFT data with the PBE exchange-correlation functional and D2 dispersion correction. Therefore a well-performing MLIP model should reproduce the density and diffusivity of that reference method rather than experimentally-correct values. This distinction is seen clearly in our density data: the 500\,K temperature which we have run our tests is likely above the experimental boiling point of the 3:7 EC/EMC mixture being studied. However, the PBE-D2 functional predicts a stable liquid under these conditions, so we consider this behaviour to be ``correct'' for the purposes of assessing model performance. 

Measuring reliable density distributions and diffusivities at the PBE-D2 level was impractical due to the high computational cost of ab initio MD, so the final GAP model of ref.~\cite{magduau2023machine},
 (labelled as Gen16/DTS in that paper) was used as a surrogate to obtain these quantities. Ref.~\cite{magduau2023machine} contains detailed comparison against both AIMD and experimental data to confirm that this model captures the correct behaviour of the liquid density, but DFT data were not available to validate the GAP-predicted diffusivity. This point will be discussed further in sec.~\ref{sec:gen0results}.

Taking all these considerations together, we suggest the following metric, $Q$ to evaluate model performance. All quantities in the following expression will be averaged across a committee of five models, to limit the effect of stochastic training on our conclusions. 

\begin{multline}
Q = {\rm max}\left(1,\frac{t_{\rm stab,NPT}}{1~{\rm ns}}\right)\\+H(1{\rm ns}-t_{\rm stab,NPT})~\left(1-{\rm tanh}[D_{\rm KL}(P_{\rm ML}(\rho)||P_{\rm ref,av}(\rho))-D_{\rm KL}(P_{\rm ref,i}(\rho)||P_{\rm ref,j}(\rho))]\right) \\
     +H(1{\rm ns}-t_{\rm stab,NVT})~\left[1-{\rm tanh}\left(\frac{|D_{\rm ML}-D_{\rm ref}|}{D_{\rm ref}}\right)\right]
     \label{eq:Metric}
\end{multline}

The three terms of eq.~\ref{eq:Metric} assign a score on [0,1] for the stability, density, and diffusivity of the MLIP respectively. Each term will be discussed and defined in detail below. Note that $Q$ is strictly a function of temperature and composition, but for computational convenience we only consider 500K trajectories at 3:7 EC:EMC. We anticipate that the general trends observed here will carry over to other conditions.

The first term of eq.~\ref{eq:Metric} scales $t_{\rm stab,NPT}$ onto [0,1]. A value of 1 is awarded to any trajectory that is stable for longer than 1ns, since we observe that trajectories which can run stably for 1ns typically remain stable indefinitely. In the second and third terms, the Heaviside step function $H(t)$ is used to ensure that only members of the MLIP committee that are sufficiently stable to compute density and diffusivity reliably are included in the quality metric.

To compare the predicted densities of our MLIPs, it is desirable to use the entire probability distribution of instantaneous densities, $P_{\rm ML}(\rho)$, sampled from an isothermal-isobaric trajectory. This quantity is much more informative than the average value $\bar{\rho}$, as it depends on the model's accuracy in both high- and low-probability regions of phase space, as well as including information on the isothermal compressibility of the liquid. We use the Kullback-Liebler divergence $D_{\rm KL}(P||Q) = \sum_{x} P(x) {\rm log}\frac{P(x)}{Q(x)}$ to quantify the dissimilarity of two probability distributions. 

Clearly $D_{\rm KL}=0$ when $P=Q$, but since this limit will never be achieved in practise we require an estimate of the ``acceptable'' divergence for two models that are considered similar. Since our reference method is itself an MLIP (the GAP model) and thus stochastic, we obtain this estimate by computing $D_{\rm KL}(P_{\rm ref,i}(\rho)||P_{\rm ref,j}(\rho))$, the divergence between two GAP MLIPs that differ only in the random seed used to sparsify their kernel matrix. $P_{\rm ref,av}(\rho)$ is the distribution averaged across a committee of reference GAP models. The tanh activation function is used as a switch: when the divergence of the target MLIP being tested from the (averaged) reference distribution is greater than that between two reference models, the tanh function becomes positive and the value of the second term decreases. Thus, an MLIP with a small $D_{KL}$ vs the reference will have a second term of 1, and a very large divergence will give a second term of 0.

In the third term of the equation, $D_{\rm ML}$ and $D_{\rm ref}$ are center-of-mass diffusion coefficients evaluated using the Einstein formula $D_\mu = <|\mathbf{r}_\mu(t)-\mathbf{r}_\mu(0)|^2>/6t$ where $\mathbf{r}_\mu$ is the centre-of-mass position for a particular molecule of type $\mu$. In each case, we evaluate this formula for a 1ns canonical trajectory at the GAP-equilibrated mean density of 0.92 g/cm$^3$. In a multicomponent system, $D_{\rm ML}$ and $D_{\rm ref}$ differ for different components; in our case we use the weighted average of the diffusivities for EC and EMC when evaluating equation \eqref{eq:Metric}, to obtain a single number for comparison. In practise, we find that EC and EMC have nearly identical diffusivities in most models. The form of the last term in eq.~\ref{eq:Metric} contributes 1 to $Q(T)$ when the fractional error in $D_{\rm ML}$ is small, and 0 when it is large.

In total, therefore, values of $Q$ less than 1 indicate a model that does not provide stable liquid-state dynamics at atmospheric pressure, values between 1 and 3 indicate stable models with increasingly good descriptions of the density distribution and diffusion coefficients of the liquid. We suggest that this metric provides a fairly complete and transferable approach to comparing diverse MLIPs (or indeed traditional forcefields). 

\subsection{DeePMD Active Learning protocol}

We devised a simple active learning procedure representative of standard practise in the literature, in which an initial model generates new configurations and committee disagreement (the variance in predicted energies/forces between a set of 5 equivalent models) is used to identify which are worst-represented in the original training data. These configurations are added to the training set, and training is repeated. Note that since DFT calculations must be used on these configurations, they are restricted to small sizes. The protocol used is as follows:

\begin{itemize}
    \item Train a set of 5 models on the previous ``generation'' of dataset. One of these models is arbitrarily designated as the primary model. We chose to restart training from new randomly-assigned network weights at each generation, to avoid trapping our models in local minima of the optimisation landscape, but initialising the training from the previous model's weights is also a valid strategy.
    \item Simulate 25 NPT trajectories with the primary model: 5 different compositions (all containing 12 molecules, going from 17\% EC to 83\% EC) each with 5 different initial configurations sampled from a GAP MD trajectory. 
    Run up to 500ps, stopping early if a high-error configuration is found (see next point), if any intramolecular bond length exceeds its average value by more than 100\%, or if the liquid density leaves the range $[0.2{\rm g/cm}^3,1.2{\rm g/cm}^3]$. 
    These last two conditions ensure that we do not select new training configurations that are drastically unphysical, since we found empirically that such configurations can damage model performance. Note that the values of these parameters still allows for retraining on non-equilibrium conditions (e.g. stretched bonds and low-density regions) in order to prevent the models from leaving the equilibrium configuration space.
    \item Select the first configuration from each trajectory that has a maximum committee variance in the force greater than 0.5 ${\rm eV}^2/{\rm Ang}^2$. This corresponds to 10x the typical training error of 10meV/Ang. 
    \item If no valid retraining configuration were selected after 500ps (or after the break criteria were hit), we simply select the configuration with the largest variance in predicted force.
    \item Recalculate energies of the 25 selected configurations with the PBE-D2 functional using CASTEP, and add them to the training set for the next generation.
\end{itemize}

\section{Results}
\label{sec:results}

\subsection{Performance of transferred datasets without active learning}
\label{sec:gen0results}

One simple approach to assessing the information content of a dataset, and hence its transferability between algorithms, is to measure the performance of a model trained only on that dataset itself (without any further algorithm-specific active-learned configurations). In this section, we present the performance of MLIPs trained using DeePMD and MACE on each component of the GAP dataset. 

\begin{figure}
	\centering
	\begin{subfigure}[t]{0.48\textwidth}	
    	\subcaptionbox{}{\includegraphics[width=\textwidth]{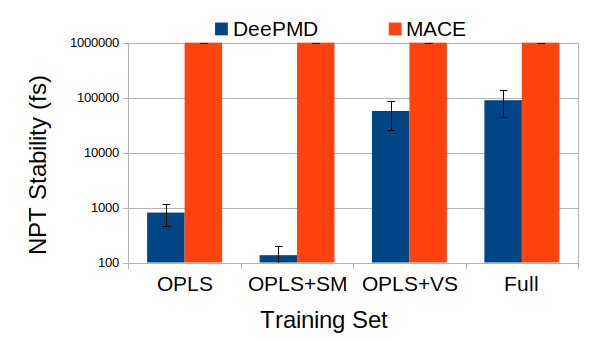}}
    \end{subfigure}	
    \begin{subfigure}[t]{0.48\textwidth}
	\subcaptionbox{}{\includegraphics[width=\textwidth]{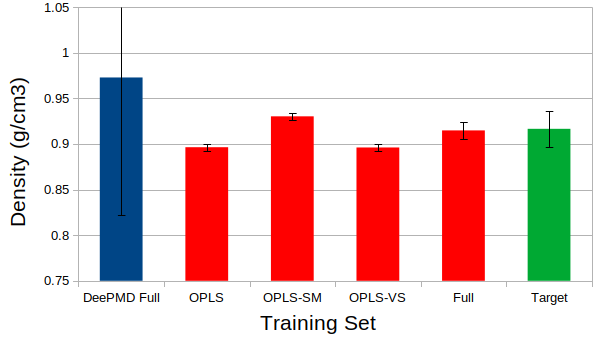}\label{fig:gen0densities}}
	\end{subfigure}	
	\begin{subfigure}[t]{0.48\textwidth}	
    	\subcaptionbox{}{\includegraphics[width=\textwidth]{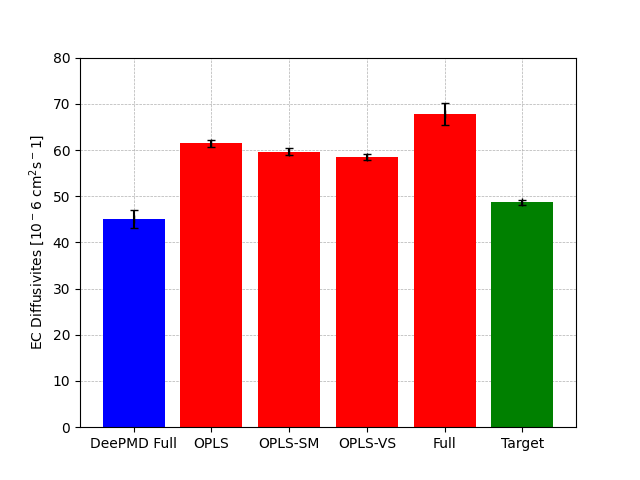}}
    \end{subfigure}	
    \begin{subfigure}[t]{0.48\textwidth}  
	\subcaptionbox{}{\includegraphics[width=\textwidth]{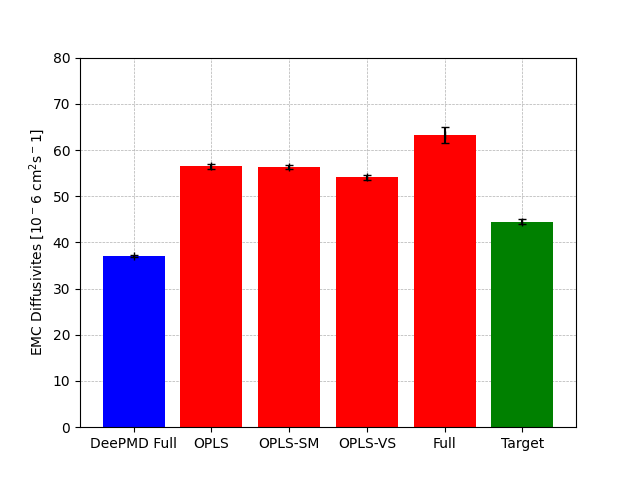}}
	\end{subfigure}
	\caption{Measures of performance for models trained on transferred data alone. All MD simulations performed at 500K and 3:7 EC:EMC compositions with 640 atoms. a) Model stabilities for naively transferred datasets under NPT conditions, comparing DeePMD and MACE models. Error bars represent a standard error across 5 equivalent models. b) Predicted mean densities for selected models, labelled according to their training dataset. Only models whose trajectories were stable for 1ns are included in the density and error analysis. 100ps of trajectory were discarded for equilibration. The green bar labelled Target represents the density of the reference GAP model\cite{magduau2023machine}, which has been validated against reference DFT calculations. Bottom: Predicted canonical-ensemble diffusivities of EC (c) and EMC (d) for selected models trained on the naively-transferred datasets. Diffusivity is computed from a linear fit to the centre-of-mass mean square displacement (MSD). The MSD was averaged over an ensmble of up to 5 models, and over a set of overlapping blocks of 100 ps with starting points separated by 10 ps. An equilibration region of 100 ps was removed at the beginning of each trajectory and a ballistic regime of 10 ps was removed from the MSD blocks prior to the diffusivity fits. The error bars represent the standard error, which was calculated by dividing the standard deviation of all diffusivity values by the square root of the number of samples i.e. total number of blocks. Only trajectories that contained at least one stable MSD block were included in the diffusivity calculations.} 
    \label{fig:gen0performance}
\end{figure}

Fig.~\ref{fig:gen0performance} shows model performance using the three key properties identified in sec.\ref{sec:metric}. We focus initially on the stability of the different models, which as argued previously is highly significant in determining the ease of initiating active learning protocols.

The first key result is that MACE models are clearly more stable than DeePMD under these simulation conditions, rarely suffering trajectory failure for any reason. We attribute this stability to the longer effective interaction range and greater smoothness of the MACE potential, both of which reduce the probability of encountering unphysical conditions and leaving the well-trained region of configuration space during a typical MD simulation. 

For DeePMD, by contrast, OPLS data alone are completely insufficient to yield a stable model liquid: the trajectories for all DeePMD models halt after a small number of timesteps, typically in a catastrophic failure where molecules break apart and energy conservation is lost. 
It appears that these models place less penalty on bond-breaking and close approach of intermolecular atoms than do the smoother and longer-ranged MACE models.

Despite breakdown of intramolecular geometry being the dominant failure mode, augmenting the training dataset by including single-molecule configurations (that explicitly sample intramolecular coordinates) provides no significant improvement to the stability of DeePMD models.
By contrast, incorporating the volume scan configurations to the training set increases $t_{\rm stab}$ by two orders of magnitude for the DeePMD models. Taken together, these results suggest that the principle cause of trajectory failure in our DeePMD models is when two molecules approach to unphysically close distances, leaving the well-trained region of configuration space and thus facilitating unrealistic breakdown of the molecular geometry. 
Volume scan configurations that sample local densities up to 2 g/cm$^3$ can teach the model not to undergo this close approach, thus greatly reducing the failure rate. This result is particularly valuable, since volume scans are a cheap and efficient way to augment a training set (with very few additional reference calculations required) thus offering a rapid route to stable initial models and reducing the requirement for iterative training. 

The main difference between the Full and OPLS+VS datasets is the inclusion in Full of active-learned configurations, i.e. configurations selected to correct deficiencies in the GAP model, so comparing these models should indicate the extent of transferability for those configurations. Note however that the additional configurations do span a range of EC:EMC compositions, so we cannot completely separate the effect of active-learned configurations from the effect of incorporating multiple compositions into the training set.

DeePMD models trained on the Full dataset are only slightly more stable than those trained on OPLS+VS: the mean stability time increases by around 40\%, comparable to the statistical error in $t_{\rm stab}$. Strictly, we cannot compare the densities of the Full and OPLS+VS models since the latter do not contain many models with $t_{\rm stab}=1$ns, however if we consider all models with $t_{\rm stab}>100$ps then both training sets yield a predicted density of 1.02g/cm$^3$, slightly higher than the reference value. 
A difference between these two datasets does emerge when we consider the diffusivities: the diffusivity predicted by the Full models (39$\times 10^{-6}$cm$^2$/s) is much closer to the reference value (48$\times 10^{-6}$cm$^2$/s) than that predicted by OPLS+VS (FuLL: 23$\times 10^{-6}$cm$^2$/s. 

These results suggest that active-learned configurations produced by one MLIP may provide little benefit to other MLIP algorithms, except possibly by reducing random force errors sufficiently to improve the predicted diffusivity. 
The fact that stability hardly improves on going from OPLS+VS to the Full dataset indicates that the high-error configurations leading to failure in the DeePMD models are qualitatively different (or at least distant in feature space) from those in the GAP model. This idea will be tested further in subsequent sections.

A similar conclusion is hard to draw for the MACE models since the stability is high for all training sets. This robustness is a key advantage of the MACE model, suggesting that classical forcefield data alone are sufficient to circumvent many iterations of active-learning when using MACE.

Turning now to the densities (fig.~\ref{fig:gen0performance}b) we see that MACE model densities are more variable than their stabilities, but that all the training sets considered yield a mean density that is roughly within error of the target (GAP) value. Again, the many-body descriptors and extreme smoothness of the MACE potential permits accurate property prediction from training data that are not only limited but also statistically different from the DFT structural ensemble. Our best-performing DeePMD models do achieve a similar performance, even without active learning, but their lower stabilities entail much larger error bars and it is difficult to assess whether these models are truly agreeing with the reference method.

A plausible interpretation of these data is that our DeePMD models are overfitting: all the training sets considered are too small to fit a neural network effectively, so the models give small force errors in well-trained regions of configuration space but have many ``holes'' where the potential is underdetermined and the errors are consequently high. MACE, which has a much lower threshold for required training data, does not suffer from this problem.

Although all MACE predicted densities are within the error bar of the reference value, those from the Full models are statistically different from the other datasets, and are remarkably close to the reference. The implication is that MACE models are further improved by including GAP active-learned configurations and multiple molecular compositions, achieving a very good-quality model without any additional iterative training. The benefit of these configurations to the DeePMD models is less clear, suggesting that the increased range of MACE makes it more able to re-use training data than the conventional neural network approach. 

Panels c) and d) of fig.~\ref{fig:gen0performance} present canonical diffusion coefficients calculated for the centre-of-mass coordinates of EC and EMC molecules, respectively. We show only those models that were sufficiently stable to compute diffusivities: DeePMD trained on the Full dataset, and all the MACE models.
Strikingly, we see that the DeePMD model gives only a slight underprediction of the diffusivity relative to the reference value, while the MACE models exceed the reference by around 50\%. The consistency of this difference across all four datasets considered suggests that this increase is a systematic effect of the MACE potential form, but it is not clear whether the MACE diffusivity is more or less accurate than the DeePMD and reference models. As mentioned previously, the reference value was obtained from the GAP model described in ref.~\cite{magduau2023machine},
 which was benchmarked against experimental and DFT density data but not against diffusivities. 
Given its superior performance in most other tests to date, and typically lower validation errors than GAP, we hypothesise that the MACE values presented in fig.~\ref{fig:gen0performance} are actually closer to the DFT diffusivities than the GAP reference. 

We have attempted to confirm this expectation by evaluating prediction errors in the forces and energies of configurations explored by the NVT trajectories of the respective models. This test is more instructive than simply comparing prediction errors for the validation set in training, since it incorporates a measure of how far each model strays from its well-trained configuration space, as well as the raw prediction error.
In fig.\ref{fig:SI_RMSE_NVT} we show the energy and force errors as a function of time along the trajectory for a well-performing DeePMD model (i.e.~one that gives density and diffusivity closely agreed to the GAP reference) and for a typical MACE model (properties were broadly similar for all models studied). We find that the RMS errors in predicted forces and energies are both small and constant along the trajectory for MACE. For DeePMD, the errors are several times larger and the magnitude of both increases substantially over the first 100ps of simulation time, implying that the model is sampling configurations dissimilar to its training data. Therefore, we conclude that the MACE models are performing more reliably and suffering smaller random errors than DeePMD and hence probably more accurate than GAP as well. The uniformity of prediction errors across phase space observed for MACE is especially relevant, since diffusivity is related to a time-correlation function.

Finally, we note that of these authors will soon publish a separate study of the EC:EMC liquid, which will show that GAP diffusivities at 300K are an order of magnitude lower than experimental values, while MACE models achieve reasonable quantitative agreement. Those data provide further indirect evidence that the MACE models here may be closer to the DFT behaviour than the GAP reference.

Our diffusivity analysis therefore leads to the surprising conclusion that MACE models trained on a small classical dataset provide better dynamic properties than GAP models with a carefully crafted iterative training dataset. This difference is probably because the high-energy barrier regions on which diffusivity depends are difficult to capture in a training dataset without careful targeted sampling. Therefore a model with good extrapolative ability and a simple dataset, like MACE, is able to outperform poorly-extrapolating models trained on complex equilibrium datasets. This finding has important implications for foundational models: MACE extrapolates well locally from little data, and has a large capacity to fit very diverse data. These properties make it suitable for developing foundational models.

\subsection{Active Learning analysis of transferability}
\label{sec:ALresults}

Notwithstanding the considerable success of the MACE models even for very small training sets, it is instructive to consider how much active-learning effort may be saved by transferring training configurations. We therefore consider how the different DeePMD models (that initially perform quite poorly) improve during the active learning procedure - specifically, whether transferring different data from the GAP training set reduces the number of active-learning iterations required to achieve acceptable model performance. 

\begin{figure}
    \centering
	\begin{subfigure}[t]{0.48\textwidth}	
    \subcaptionbox{}{\includegraphics[width=\textwidth]{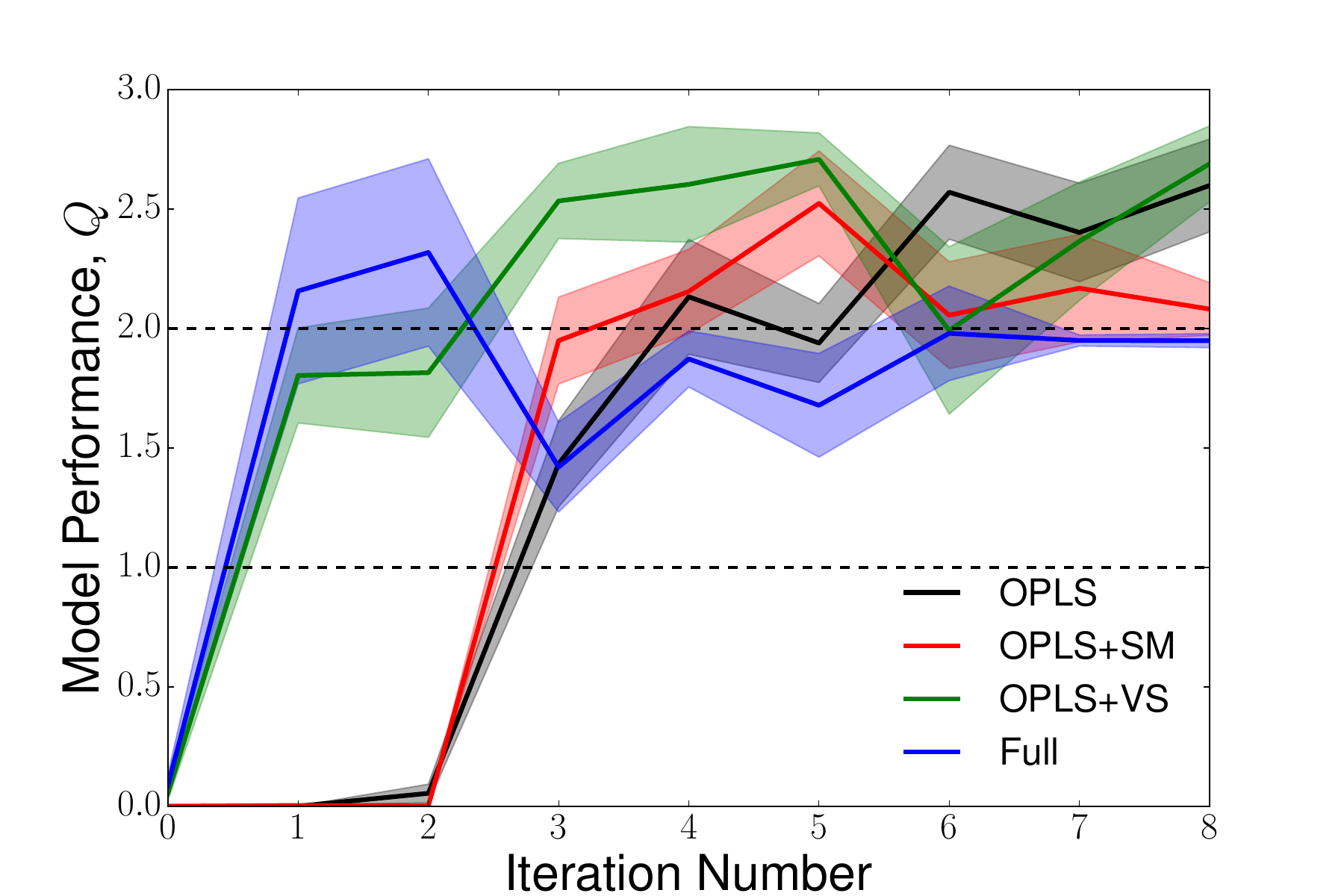}}  
    \end{subfigure}	
    \hfill
	\begin{subfigure}[t]{0.48\textwidth}	
    \subcaptionbox{}{\includegraphics[width=\textwidth]{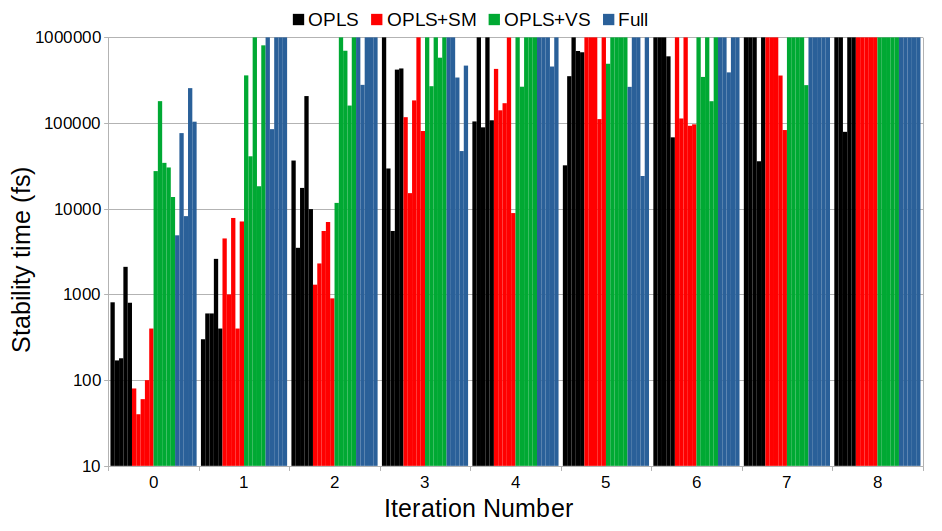}}
    \end{subfigure}

	\begin{subfigure}[t]{0.48\textwidth}	
    \subcaptionbox{}{\includegraphics[width=\textwidth]{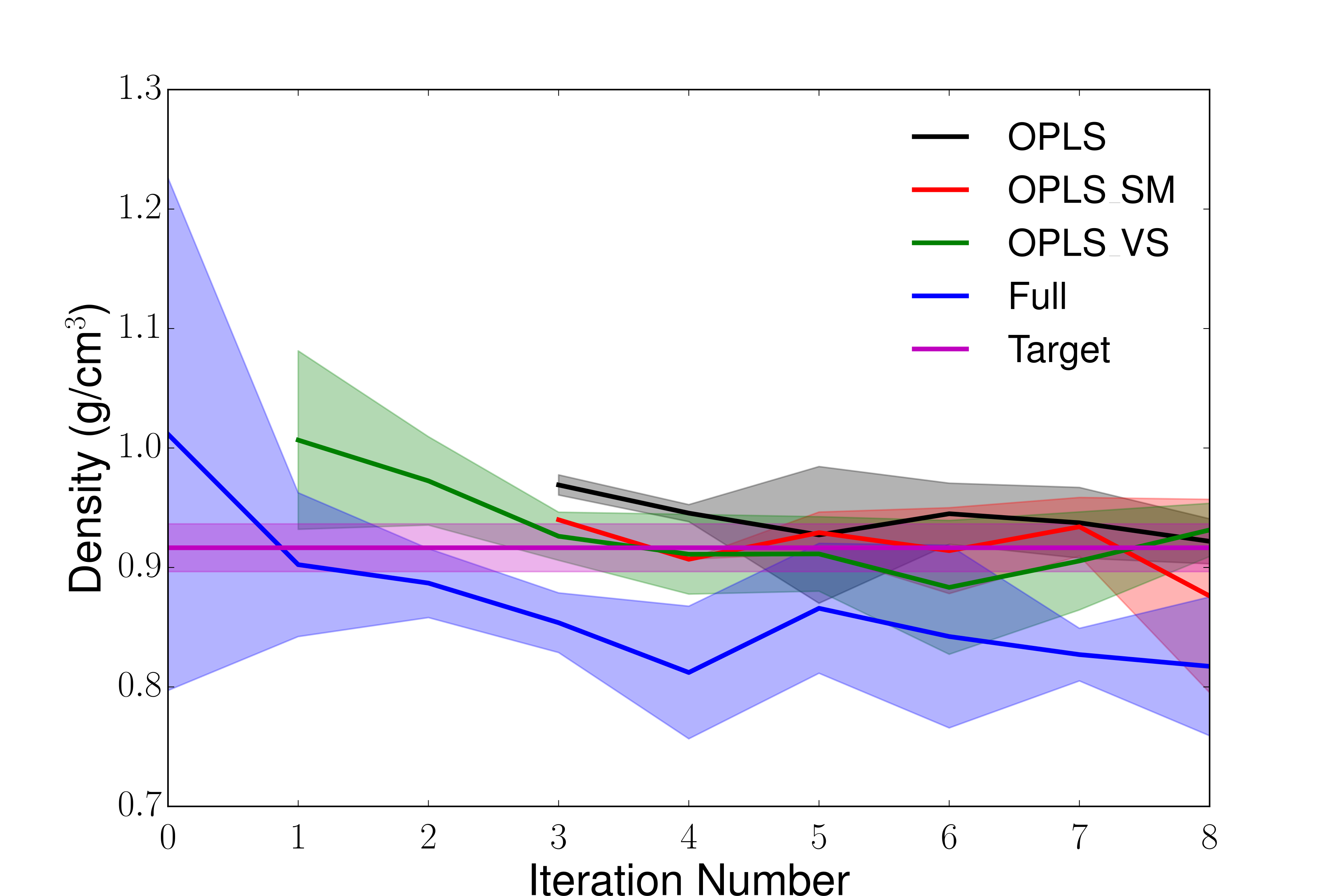}}
    \end{subfigure}	
    \hfill       
    \centering
	\begin{subfigure}[t]{0.48\textwidth}	
    \subcaptionbox{}{\includegraphics[width=\textwidth]{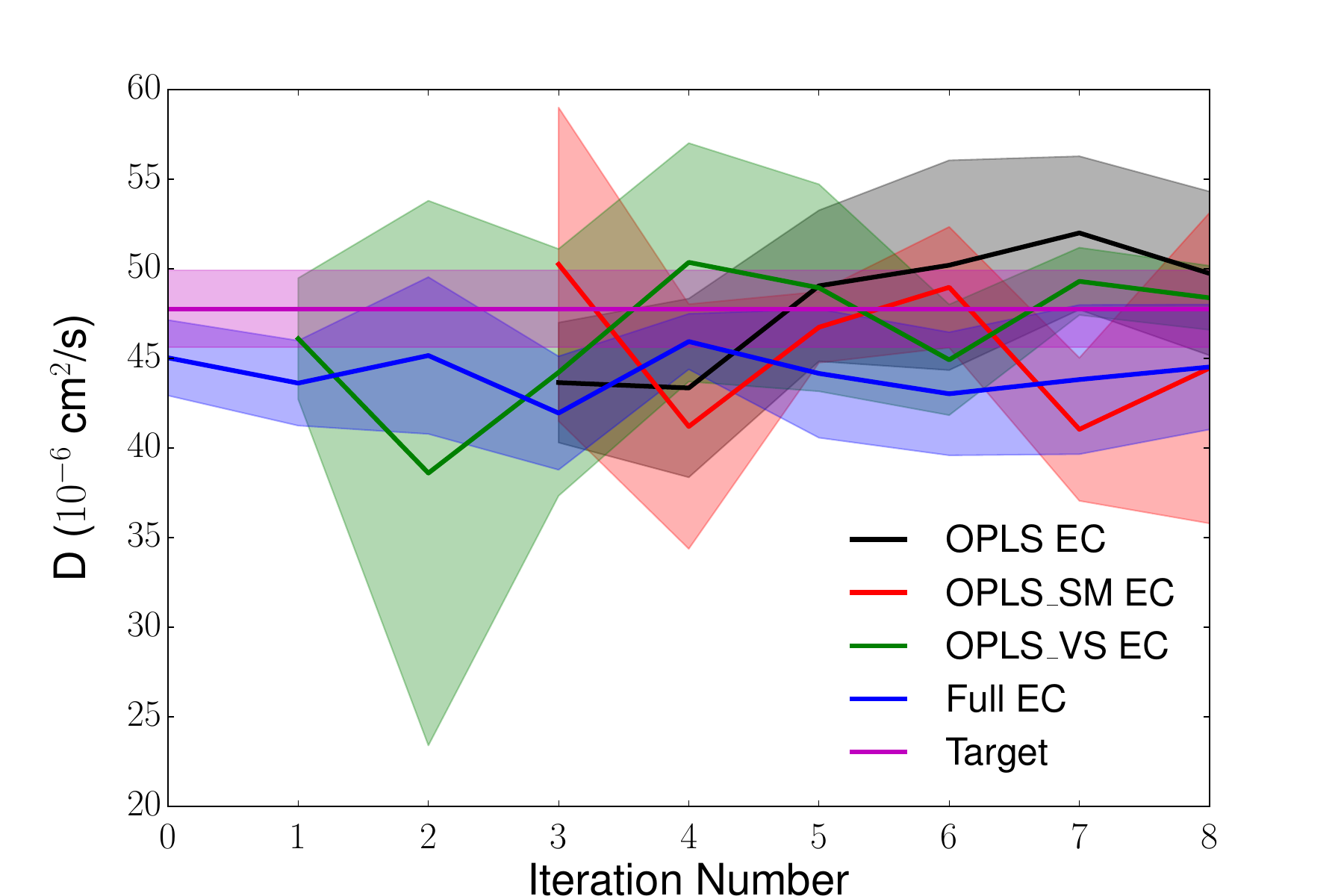}}
    \end{subfigure}	
 
    \caption{Performance of DeePMD models starting from different datasets as a function of active learning iterations. Panel a) Overall model quality, as defined in eq.~\ref{eq:Metric}, as a function of iterative training. 
    b) Stability times (time until trajectory evaporates or molecules break irreversibly) as a function of active learning generation for different initial training sets. Each column represents a different MLIP model, all models with the same colour are trained on identical training sets with different random seeds. c) Mean density as a function of active learning iterations. The solid lines represent the mean density of a trajectory, 
    averaged across 1-5 models - all models capable of simulating 1ns of stable trajectory are included in the average. The shaded regions indicate the standard error of the average density, computed across the set of stable models. d) Mean diffusivity as a function of active learning iterations. The lines and areas have the same meaning as panel c), except that here data were collected from NVT trajectories.}
     \label{fig:ALperformance}
\end{figure}

Fig.~\ref{fig:ALperformance} summarises how each measure of DeePMD model performance changes with number of active-learning iterations, starting from each of the four initial training sets. Thus, the generation-0 models for each training set correspond to the models described in the previous section.

Panel a) presents the model performance metric introduced as eq.~\ref{eq:Metric}. We see immediately that successive generations of active learning improve the performance of all models. A particularly rapid increase occurs after $t_{\rm stab}$ reaches 1 (between iteration 1 and 3 depending on the model). This transition is largely because models with $t_{\rm stab}$<1ns may already have densities and diffusivities approaching the target values, but which do not contribute to $Q$ until the Heaviside function condition in eq.\ref{eq:Metric} is satisfied. Our metric is therefore quite susceptible to the transition from ``unstable'' to ``stable''. Nevertheless, this choice of metric serves to emphasise that early iterations of active learning are mostly spent training the models to avoid pathological failures, and only once these ``holes'' are corrected do active-learned configurations begin to correct the predicted densities and diffusivities.

The main result of panel a) is that commencing active-learning with more transferred data decreases the number of iterations required to reach reasonable performance - including configurations generated by GAP active-learning. 
However, all four models achieve a similar performance after about 6 iterations of active learning and $Q$ approaches a plateau, suggesting that achieving complete convergence to the reference values is mostly controlled by incorporation of algorithm-specific active-learned configurations and/or multiple local compositions. 
We also note that the Full dataset models reach an earlier and slightly lower $Q$ plateau than the other models - largely because the predicted density of this model converges to a value slightly below the reference. This deficiency will be discussed further in the next section.
As in the previous section, incorporating the volume scan configurations yields the greatest improvement to DeePMD model performance, with little additional benefit (or even a penalty) arising from the added active-learned configurations in the Full models.

Panel b) shows that the improvement of model stability with active-learning generation mirrors the trend in stabilities at generation 0: the models with the most transferred data (and hence the highest initial stability) quickly improve, reaching $t_{\rm stab}\approx 1$ns after 2-3 generations. The data-poor models (OPLS and OPLS+SM) take significantly longer to achieve this stability - typically 6-8 generations. This difference is largely because trajectories driven by the more stable models explore closer to the correct configuration space, hence the configurations they generate for the active-learning protocol are more relevant to subsequent dynamics. Configurations collected from short, unstable trajectories with the OPLS-based models are likely to be far from the desired equilibrium space and confer only moderate improvements to stability.

Each class of model initially predicts a mean density substantially different from the target (typically over 10\% higher).
This difference shrinks over successive generations, usually accompanied by a reduction in the standard error of the predicted quantity because more models are stable enough to make a prediction. Nevertheless, we observe that model families that start from the Full dataset reach agreement with the target after only 1-2 generations, while families starting from only OPLS data require at least 6 generations to give an accurate measure of liquid thermodynamics and kinetics alongside stable dynamics.

These results confirm the utility of transferring data, especially transferring configurations such as volume scans that will contribute significantly to improving model stability. On incorporating this additional training data, similar improvements are seen in the time required to converge predictions of liquid properties.

Interestingly, DeePMD models agree more closely with the GAP diffusivities than they do with the density, even though diffusivities depend more strongly on high-energy ``barrier'' configurations that will usually be worse-described in a training dataset, and even though the GAP values appear to be systematically underestimated (as discussed previously). 
The diffusivity of every DeePMD model considered falls within error of the target GAP value for the entire range of iterations where stable trajectories were identified. 
This agreement could indicate that force prediction errors exert less effect on the diffusivity than they do on the density, so that models in the low-data regime converge the latter property more slowly. The commonality of errors between the two MLIPs may simply indicate that their force errors have similar magnitude (while MACE is significantly more accurate for unseen configurations). Recall that our diffusivities were computed using NVT simulations at a fixed density close to the correct reference value, in order to deconvolute the error in predicted density from the errors in diffusivity that arrive from inaccurate MLIP forces. Therefore the average magnitude of the force prediction errors is probably the key quantity determining agreement or otherwise of diffusivities for different MLIPs.

\subsection{Understanding the information content of transferred datasets}
\label{sec:VolScans}

We can probe further the specific limitations of each transferred dataset by examining in more detail how the models trained on them perform. We consider the ability of transferred MLIPs to describe density variations in a controlled validation set that is constant for all models, and where the relationship between the different configurations is clear and known. We focus on naive DeePMD models without active learning, since these showed the greatest variation in accuracy.

\begin{figure}
    \includegraphics[width=0.6\textwidth]{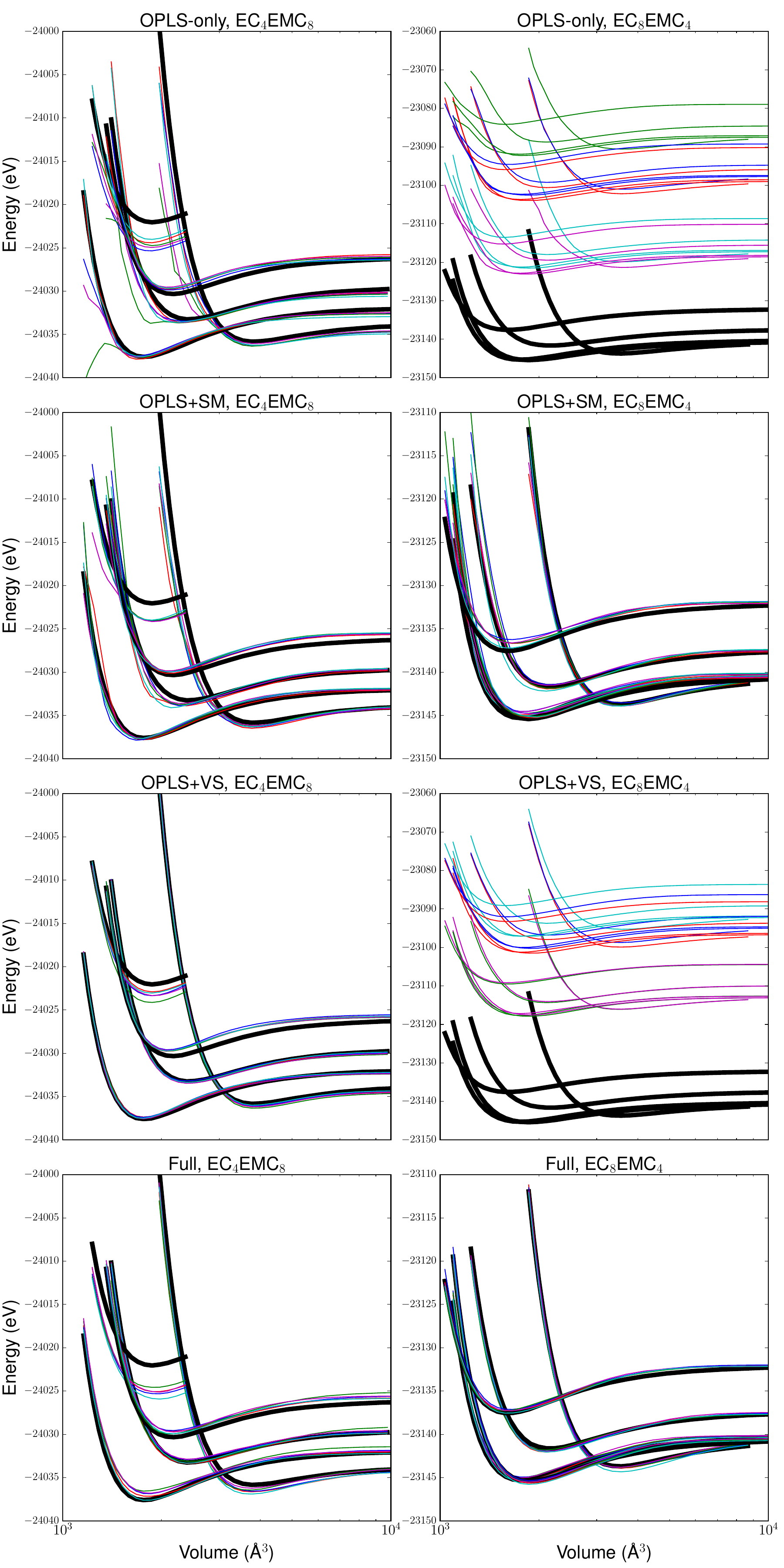}
	\caption{Potential energy prediction errors for DeePMD models trained on transferred data without active learning, evaluated for volume-scan data sets. Each curve corresponds to the energy of a particular molecular arrangement during isotropic expansion and compression. The left column of panels shows simulation cells containing 4 EC molecules and 8 EMC molecules, the right shows EC$_8$EMC$_4$ configurations. Each row represents a different transferred dataset, labelled according to the convention in sec.~\ref{sec:config_types}. 
    In each panel, thick black lines indicate the DFT energies. Each coloured line represents a different DeePMD model. For each volume scan, 5 models are shown - these models are trained on the same data but have different random seeds or slightly different learning schedules. }
	\label{fig:VolScans}
\end{figure}

Fig~\ref{fig:VolScans} shows how these models predict that the energy of a small EC/EMC cell varies under isotropic compression or expansion at constant intramolecular geometry - i.e.~a volume scan, as used to generate the VS configurations in the GAP training set. In all cases these energies diverge at small volumes due to overlapping atoms, and tend to a finite plateau in the large-volume limit where every molecule becomes isolated and non-interacting. Validating an MLIP against a volume scan tests shows both how a model describes environments at non-equilibrium intermolecular distances, and also the intramolecular energy of an isolated molecule.

We compare volume scans for 5 configurations of a 1:2 EC:EMC liquid, and 5 configurations of 2:1 EC:EMC. This choice is significant because (without active learning) only the Full training set contains configurations with compositions other than 1:2. 
Two types of prediction failure are immediately apparent from fig~\ref{fig:VolScans}. In some cases, the shape of the DeePMD energy curve changes significantly at low volumes, resulting in errors up to several eV from the DFT energy. These errors probably correlate with the ``holes'' noted in previous sections, that are responsible for catastrophic trajectory failure. In particular, note that high-density configurations for the OPLS and OPLS+VS sets often have severely under-estimated energies, permitting unphysically close approach of different molecules. These major deviations decrease significantly for larger training set sizes, with OPLS+VS and Full datasets having generally smaller RMS prediction errors in line with their improved model performance noted previously.

The second prediction error takes the form of a large shift (up to 10eV) in potential energy relative to the DFT value, consistent for all scans with a given model and roughly independent of cell volume. This error in the baseline energy (i.e. the zero-density limit) only occurs in the 2:1 composition curves, and arises because models trained on a single composition cannot learn correctly how this baseline energy should be partitioned between EC and EMC molecules. Therefore, when the composition changes, the baseline shifts by an incorrect amount. This baselining issue is potentially serious for large simulation sizes, where individual atomic environments may sample very different local compositions over time, resulting in large fluctuations in the total predicted energy which could prompt unphysically large responses from thermostats (for example).

However, the baseline error is easily corrected by incorporating isolated molecule configurations into the training set, or by training on multiple compositions as in the Full dataset. Note that training on the volume scans themselves (OPLS+VS) does not contain information about the correct partitioning of molecular energies, so the incorrect baselines persist in panel 6 of fig.~\ref{fig:VolScans}.

\begin{figure}
	\centering
	\begin{subfigure}[t]{0.45\textwidth}	
    	\subcaptionbox{}{\includegraphics[width=\textwidth]{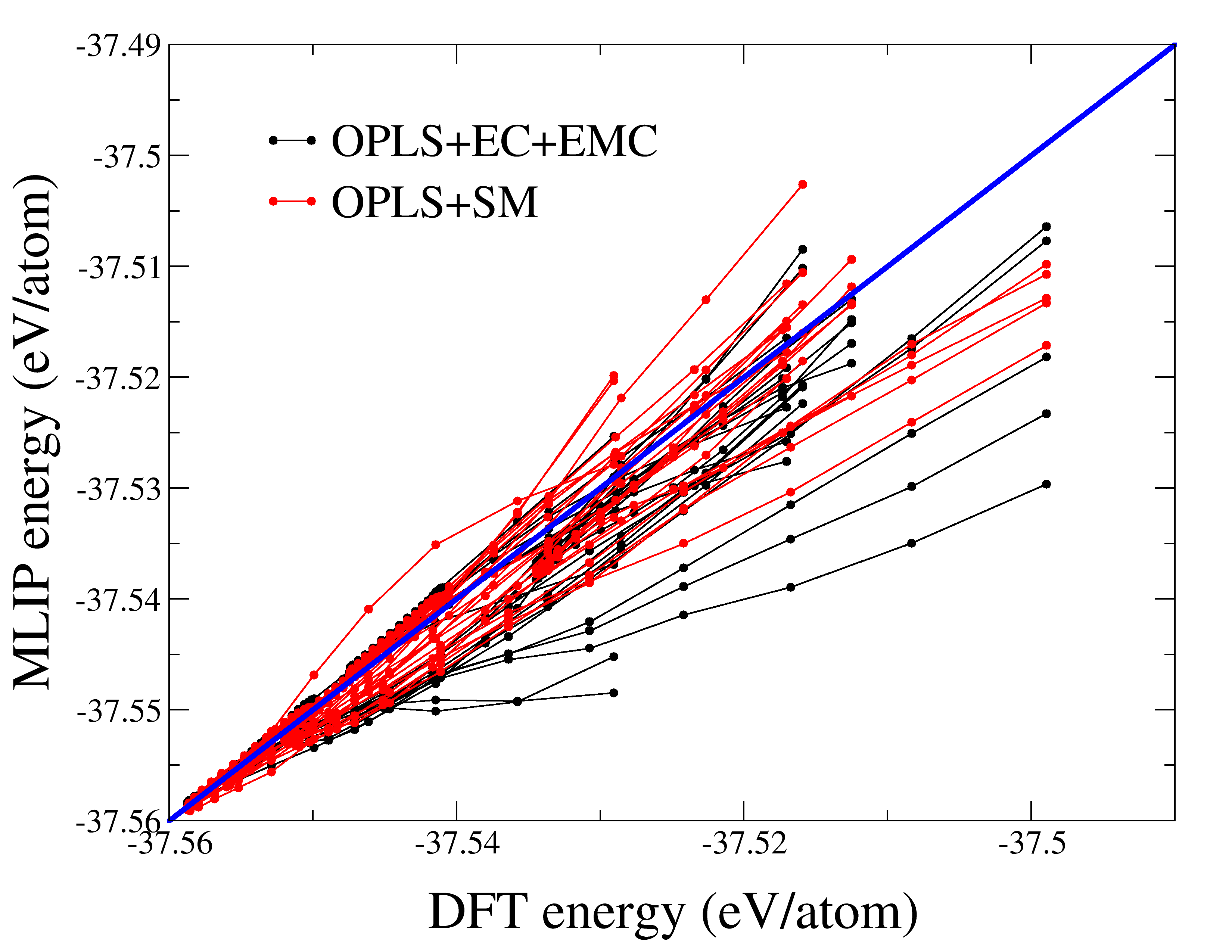}}
    \end{subfigure}	
    \begin{subfigure}[t]{0.45\textwidth}
	\subcaptionbox{}{\includegraphics[width=\textwidth]{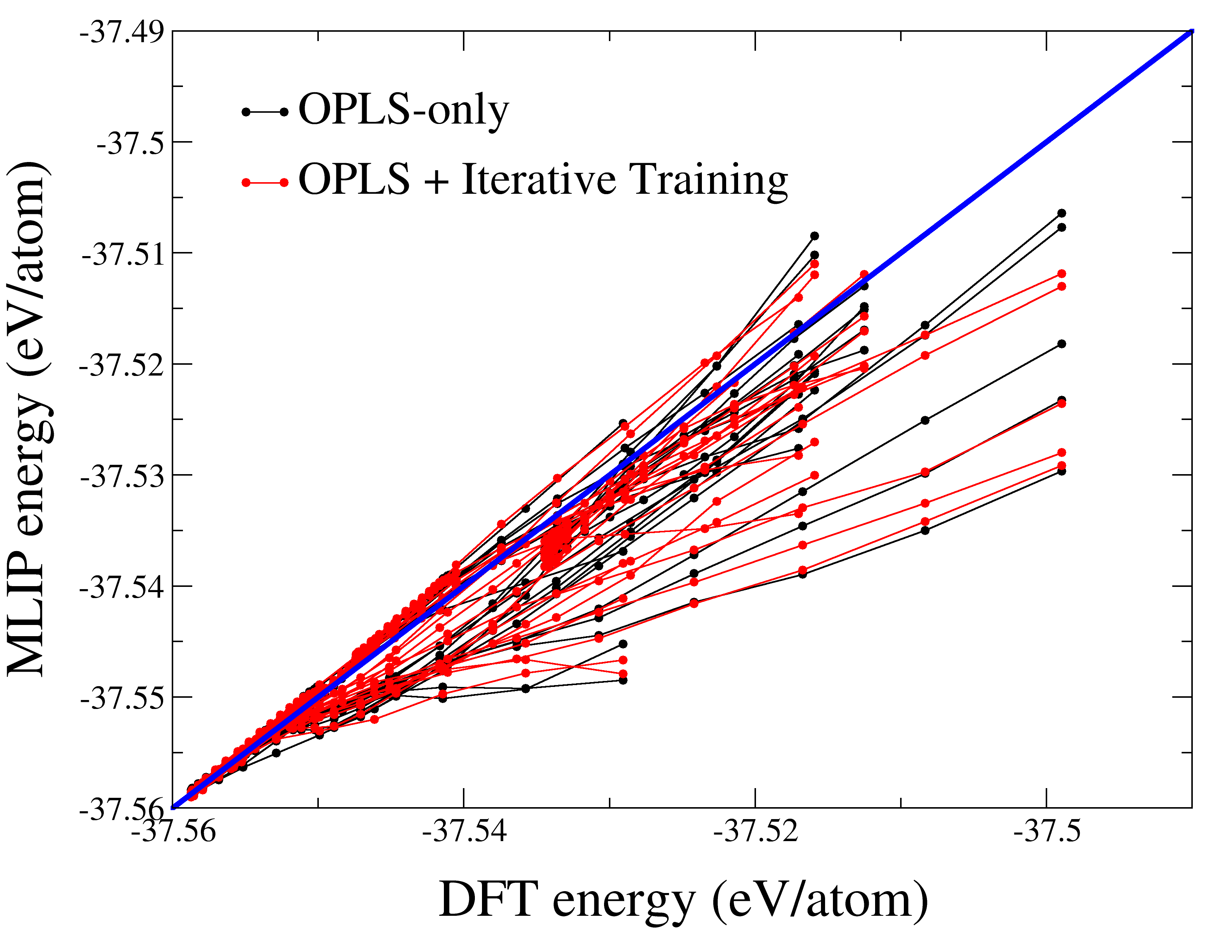}}
	\end{subfigure}
	\hfill
 
	\begin{subfigure}[t]{0.45\textwidth}	
        \subcaptionbox{}{\includegraphics[width=\textwidth]{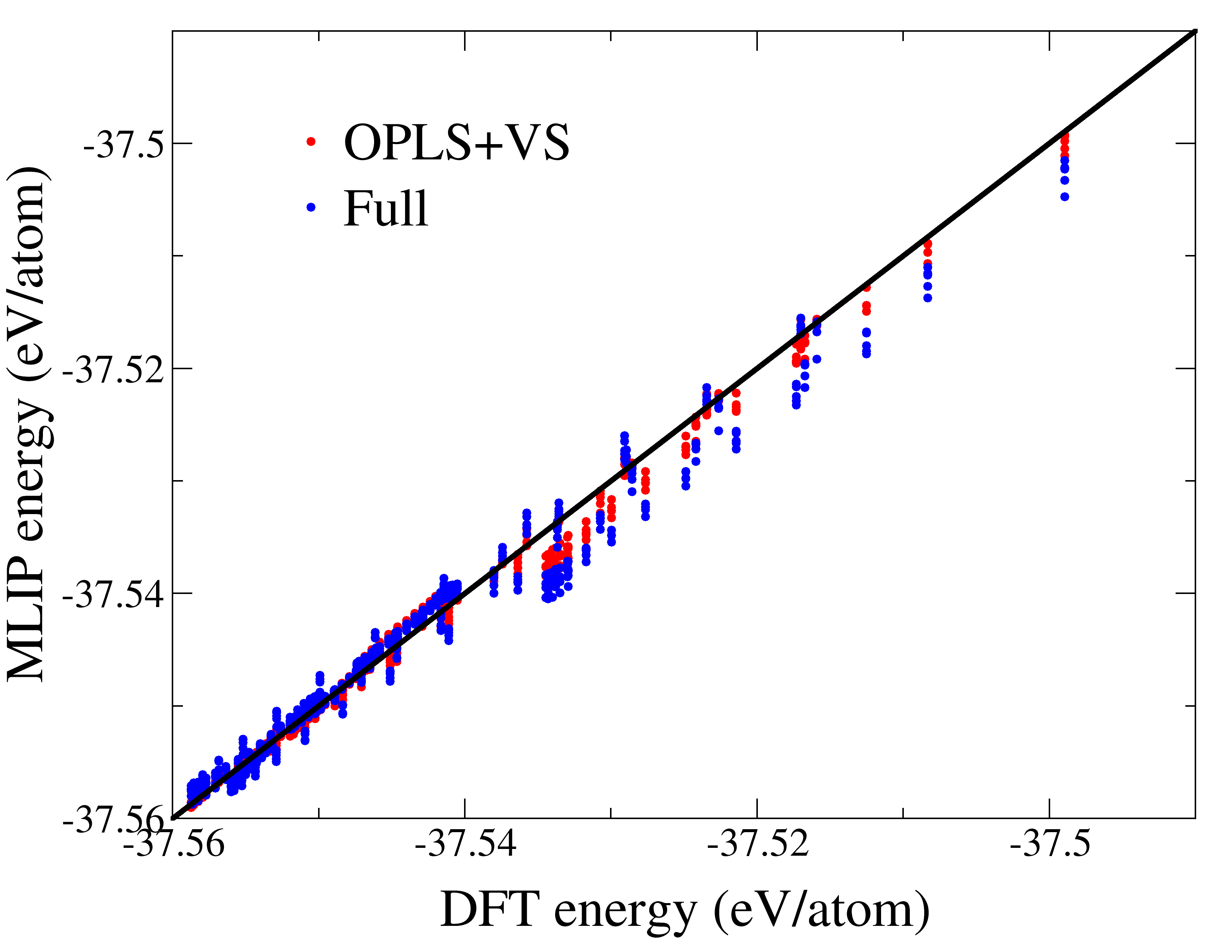}}
    \end{subfigure}	
    \begin{subfigure}[t]{0.45\textwidth}
        \subcaptionbox{}{\includegraphics[width=\textwidth]{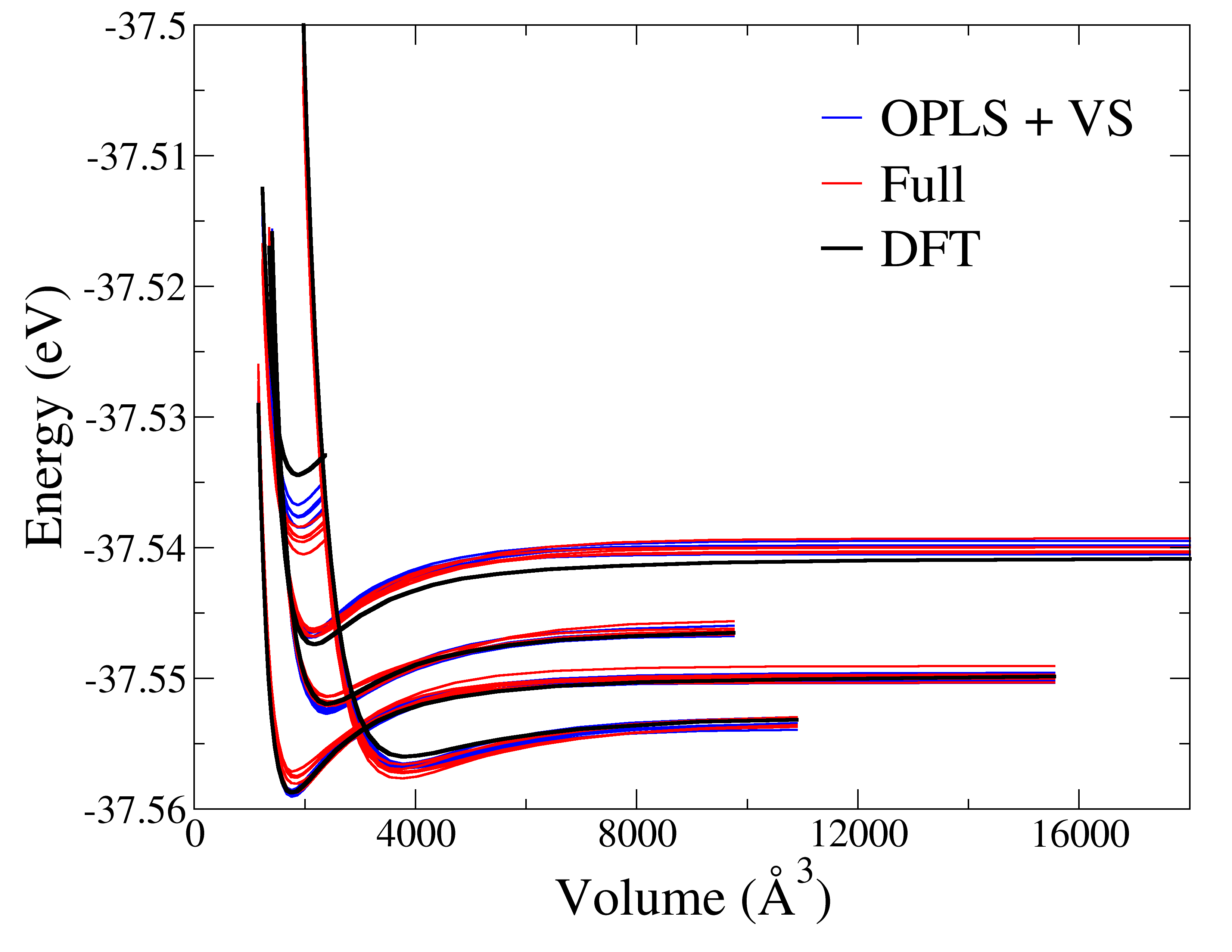}}
	\end{subfigure}
	\hfill
	 
	\caption{a): Energy prediction parity plot for volume scans of 1:2 EC:EMC, showing that OPLS+EC+EMC models (see text) give slightly poorer predictive performance than models trained with many isolated-molecule configurations, particularly at high energies (corresponding to high densities). The thick blue line indicates parity (perfect prediction). b) Parity plot comparing energy predictions for the OPLS-only training set and for a set that combines OPLS data and the subset of GAP active-learned configurations that have a composition of 1:2 EC:EMC. c) Parity plot comparing energy predictions for models trained on OPLS+VS and Full datasets. d) Energy as a function of cell volume for an ensemble of models trained on the OPLS+VS set and on the Full set, compared with DFT.}
	\label{fig:VolScans2}
\end{figure}

We found that the minimum information needed to correct the zero-density energy limit for the OPLS dataset was to add one isolated-EC configuration and one isolated-EMC, weighted so that each is included in 5\% of training steps. We refer to this training set as OPLS+EC+EMC. However, using the OPLS+SM dataset with its larger number of single-molecule configurations also corrects the baseline, and confers a 30\% reduction in the RMS error of the energy predictions compared with OPLS+EC+EMC - see fig.~\ref{fig:VolScans2}, panel a). These findings suggest that explicitly incorporating single-molecule configurations is an efficient strategy to achieve transferability between compositions, although as described in section~\ref{sec:ALresults} this approach may not substantially benefit trajectory stability and other properties.

Fig.~\ref{fig:VolScans2} highlights two further key results from the volume-scan analysis. The first is that no substantial error reduction results from the inclusion of GAP active-learned configurations in the transferred dataset without also including extra compositional data (fig.~\ref{fig:VolScans2}b). For this test, we introduce a new transferred dataset comprising the OPLS data and only those configurations of the GAP dataset that were obtained from active-learning on the target 1:2 EC:EMC composition. The parity plot shows that the errors from the model trained on this dataset are close in magnitude and in bias to the OPLS-only model. This result does not necessarily indicate that the GAP active-learned configurations offer no improvement to the model at all, since they may help to prevent MD trajectories visiting unphysical regions of configuration space, but they confer no specific advantage in describing these high- or low-density configurations.

Returning briefly to fig.~\ref{fig:VolScans}, the lowest two panels suggest that the Full models have slightly larger errors for the 1:2 EC:EMC liquid than the 2:1 liquid - in contrast to all other models where the trend is reversed. In fig.~/\ref{fig:VolScans2}c) and d), we further show that errors in 1:2 EC:EMC for the Full model are worse than those for the OPLS+VS model - particularly for the highest-energy volume scan - despite the Full model performing equally or better than OPLS+VS in most respects.

This deficiency may account for the systematic underprediction of densities in the Full model during active training (see fig.~\ref{fig:ALperformance}c), since it suggests that training sets containing multiple compositions (such as the Full models) are worse at describing the target 1:2 composition of the liquid. In this view, the lower statistical weight of the 1:2 composition in the Full training set increases the size of the phase space that the model attempts to learn, thus decreasing its fit quality in the specific composition region that we then test. If one wishes to capture a range of compositions correctly using a simple MLIP, much larger datasets are apparently required than if one is interested only in a single liquid composition. Increasing the complexity of the neural network used might ameliorate this problem, but would require even more training data to work effectively.

As a corollary to this argument, we suggest that the slightly higher stability of the Full models relative to OPLS+VS (observed both with and without active learning) arises more from their ability to describe a variety of local liquid compositions than from the inclusion of GAP active-learned configurations in this dataset. More work would be required to test this suggestion, however.

\subsection{Performance of pre-trained models}

We tested the performance of two pre-trained MACE models - MACE-MP-0\cite{MACE_MP_0} and MACE-OFF\cite{MACE_OFF} - for our EC/EMC test system, to understand the importance of system-specific training data. The former is an extremely general foundation MLIP that aims to describe atomic and molecular materials across the periodic table. It is trained on the Materials Project database, that uses DFT energy calculations with the PBE functional. MACE-OFF is designed to simulate organic and biomolecules, and is trained on a subset of the SPICE dataset with energies calculated at the $\omega$B97M-D3(BJ)/def2-TZVPPD level of theory (significantly more accurate than PBE).

The training data for MACE-MP-0 uses PBE DFT energies with no dispersion forces, hence liquid trajectories with this model evaporate rapidly. When we add a Grimme D3 dispersion correction\cite{Grimme2010} to the MLIP forces, we obtain a stable liquid with predicted density approximately 0.72g/cm$^3$. This prediction is significantly worse than the best-performing DeePMD and MACE models trained on bespoke EC:EMC data, which indicates that fine-tuning on system-specific configurations would be required to use this model practically for organic liquids. 
We note in passing that the underpredicted density and need for fine-tuning are both consistent with recent findings that foundation models are often systematically soft, with a relatively small number of high-energy configurations required to correct them.\cite{Deng24} We therefore anticipate that MACE-MP-0 fine-tuned with a small number of OPLS configurations should perform very well. 

The higher level of theory used to train MACE-OFF means that it is not directly comparable with our reference data. This MLIP model predicts evaporation of the EC:EMC liquid at around 400K, which is probably much closer to reality than the PBE-D2 models used elsewhere in this work (pure EC boils at 516K and EMC at 380K). 
The mean density of the model was 1.15g/cm$^3$, slightly different from the experimental value of 1.09g/cm$^3$,\cite{wang2022current,magduau2023machine} but we cannot at present tell whether this variance results from a model transferability error or from the functional used to train MACE-OFF.
Unsurprisingly, for a model specifically and carefully trained to describe organic liquids at a high level of theory, the MACE-OFF predictions appear to be more robust than MACE-MP-0 and more faithful to experiment than the purpose-trained MACE and DeePMD models presented earlier. MACE-OFF would likely be a suitable approach for generating initial training data for future models (even if these configurations are recalculated at a lower level of theory for computational convenience). However, it is hard to assess how well this model would perform when tested outside of its training data (e.g. for a practical electrolyte containing highly-charged ions).

\subsection{Effect of transferred data on chemical generalisability}
\label{sec:Other_molecules}

An attractive characteristic of well-trained MLIPs for molecular liquids is some degree of generalisation to different molecules, to avoid the expense and effort of training multiple models to study related chemicals (e.g.~when comparing performances of two candidate electrolytes for a particular application).\cite{dajnowicz2022high} Here, we assess whether different amounts of transferred training data have a significant impact on the generalisability of the resulting model.

\begin{figure}[h]
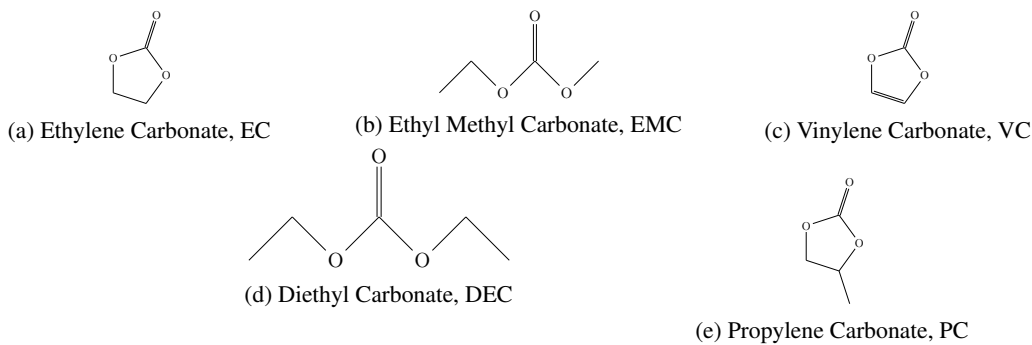

	\centering
	\begin{subfigure}[t]{0.3\textwidth}	
 \centering
    \resizebox{!}{0.22\textwidth}{\chemfig{*5(--O-(=O)-O-)}}
    \caption{Ethylene Carbonate, EC}
    \end{subfigure}	
    	\begin{subfigure}[t]{0.3\textwidth}	
 \centering
    \resizebox{!}{0.22\textwidth}{\chemfig{-[1]-[7]O-[1](=[2]O)-[7]O-[1]}}
    \caption{Ethyl Methyl Carbonate, EMC}
    \end{subfigure}	
	\begin{subfigure}[t]{0.3\textwidth}	
 \centering
    \resizebox{!}{0.22\textwidth}{\chemfig{*5(=-O-(=O)-O-)}}
    \caption{Vinylene Carbonate, VC}
    \end{subfigure}	
	\begin{subfigure}[t]{0.5\textwidth}
  \centering
    \resizebox{!}{0.18\textwidth}{\chemfig{-[1]-[7]O-[1](=[2]O)-[7]O-[1]-[7]}}
    \caption{Diethyl Carbonate, DEC}
    \end{subfigure}	
    \begin{subfigure}[t]{0.22\textwidth}
     \centering
    \resizebox{!}{0.3\textwidth}{\chemfig{*5(-(-)-O-(=O)-O-)}}
    \caption{Propylene Carbonate, PC}
	\end{subfigure}
    \caption{\label{fig:SIstructures}Structures of the molecules used in this work} 
\end{figure}

We analyse the performance of the EC/EMC DeePMD and MACE models when applied to three different chemical systems: EC/DEC 1:1 v/v, VC/EMC 1:1 v/v, and pure PC. DEC is diethyl carbonate, VC is vinylene carbonate, and PC is propylene carbonate.
Structures of these molecules are provided in fig.~\ref{fig:SIstructures}. All three liquids are commonly used in battery electrolytes, and each tests a different aspect of MLIP generalisation. EC/DEC is almost identical to EC/EMC, and is currently a popular solvent for sodium-ion batteries.\cite{DarjaziGerbaldi24} PC may be used as a single solvent or co-solvent, having greater thermal and oxidative stability than EC, but unlike EC it does not significantly passivate graphitic anodes against further reaction.\cite{FanAppMatInterfaces22}  PC is chemically similar to EC but contains a branching structure not found in the original EC/EMC training data, so this molecule serves to test the ability of an MLIP to generalise molecular shapes. VC is structurally similar to EC but chemically distinct since it possesses a carbon-carbon double bond. It is commonly used as an additive to a variety of battery solvents, since it is believed to form more stable solid-electrolyte interphase structures.\cite{VC} As an additive it is typically employed in lower concentrations than tested here, but we selected a composition where MLIP descriptions of the VC molecule would have a significant effect on the behaviour of the full liquid.

For these molecules we had no high-quality AIMD data against which to compare, and hence no reference values for the density or diffusivity. Instead our analysis focuses on stability time as previously defined, and on analysing the prediction errors against DFT for a sample of configurations drawn from the MLIP trajectories.
For each molecular system, we tested four MACE models (one for each transferred dataset) and eight DeePMD models - corresponding to generations 0 and 8 of the active-learning scheme for each transferred dataset.
Note that unlike previous sections, results here are not averaged across an ensemble of models (for reasons of computational cost). 

\begin{figure}
	\centering
	\begin{subfigure}[t]{0.45\textwidth}	
    	\subcaptionbox{}{\includegraphics[width=\textwidth]{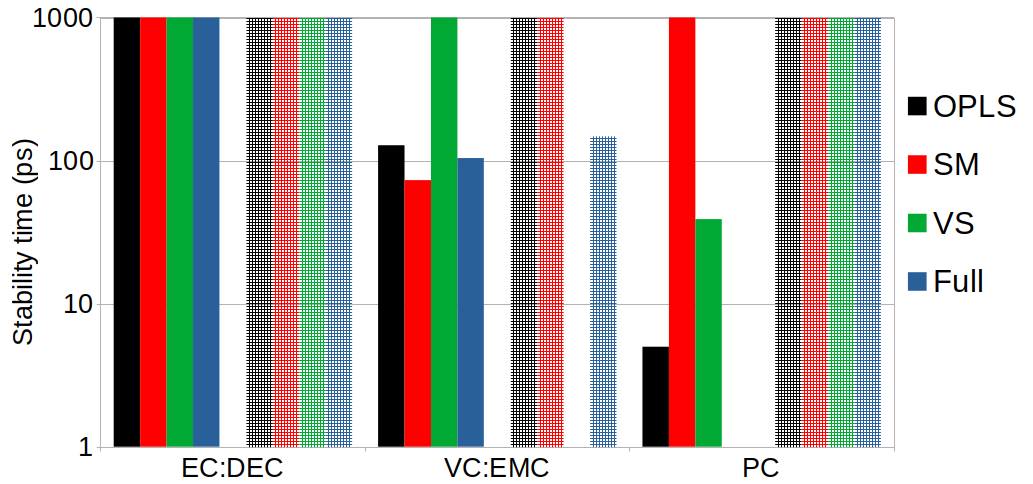}}
    \end{subfigure}	
	\begin{subfigure}[t]{0.45\textwidth}
 	  \subcaptionbox{}{\includegraphics[width=\textwidth]{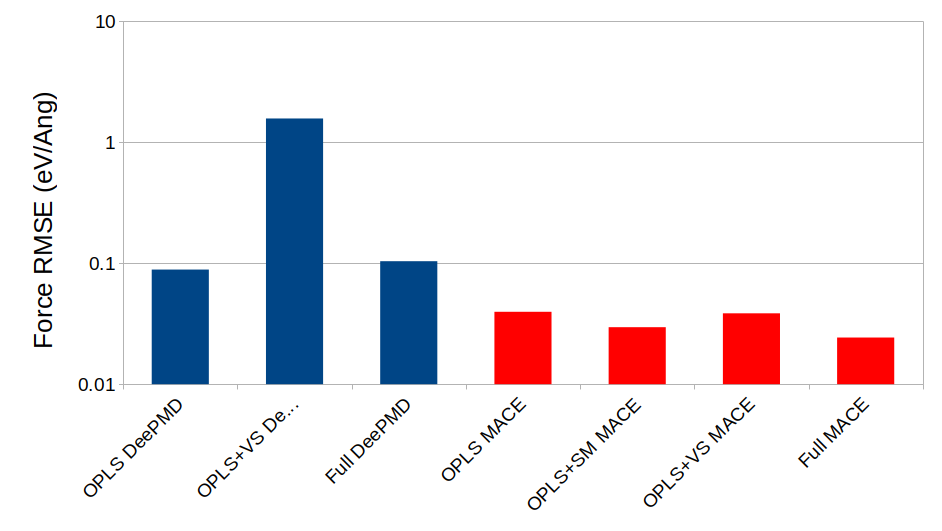}} 
    \end{subfigure}	    
    \begin{subfigure}[t]{0.8\textwidth}
    \subcaptionbox{}{\includegraphics[width=\textwidth]{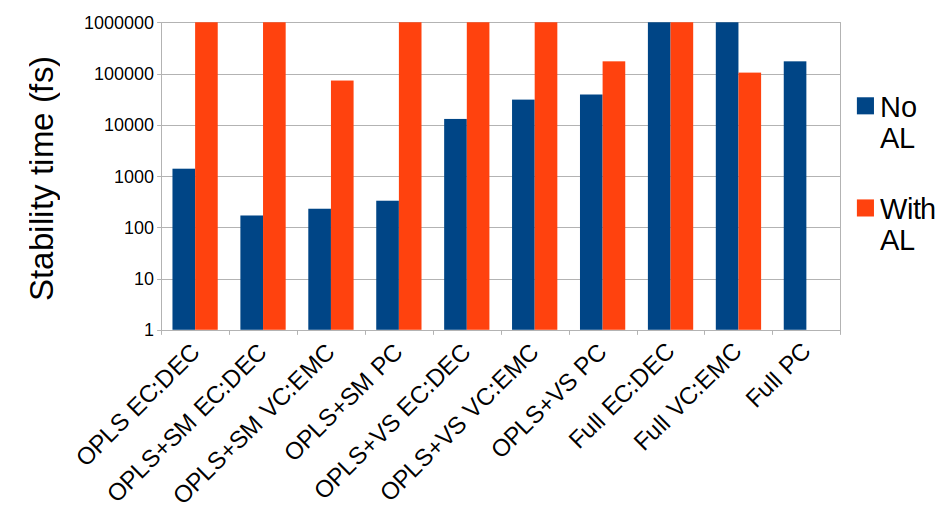}}
	\end{subfigure}
	\hfill
	 
	\caption{Comparison of different model performances when generalising to untrained chemical systems. In all panels, different initial training sets are labelled as "OPLS", "SM", "VS" or "Full" as in previous sections. a) Comparing NPT stability times for MACE and DeePMD models when generalised to a different chemical system. Solid bars represent DeePMD models with 8 generations of active learning on top of the indicated initial dataset, hatched bars represent MACE models trained on the initial dataset alone. b) RMS errors (vs DFT) in force predictions for EC:DEC liquid for selected stable MLIPs. These errors were computed over a sample of configurations from the trajectory of the model in question, rather than from a fixed test set. Blue bars indicate DeePMD models after 8 generations of active learning, red represent MACE. c) Comparing stability times for DeePMD models with and without active learning on EC:EMC. Different bars represent different model/molecule combinations.}
	\label{fig:OtherMolecules}
\end{figure}

Fig.~\ref{fig:OtherMolecules}a) shows the stability times for different molecules and models. In EC:DEC, $t_{\rm stab}$ for a given model closely mirrors its EC:EMC value. Thus, in this panel where MACE models are compared to 8th-generation DeePMD, all models are completely stable. This result is unsurprising, since there are no atom types in DEC that are not present in EMC and the atomic environments present in the two molecules probably have similar short-ranged descriptor functions.
Fig.~\ref{fig:OtherMolecules}b) shows prediction errors for atomic forces in EC:DEC, computed for a sample of configurations taken from a stable NPT trajectory. With the exception of the anomalously large error for DeePMD with the OPLS+VS training set, the errors in all models are comparable with their EC:EMC force errors (approximately 0.1\,eV/\AA~for DeePMD and 0.025\,eV/\AA~for MACE) - again suggesting good generalisation from EMC to DEC.

By contrast, VC molecules contain a functional group not present in the training sets (the alkene moiety) which our MLFFs are typically unable to describe correctly. Most models fail within 100ps for VC:EMC trajectories, typically through unphysical breaking of a C-H or C-C bond. Surprisingly, some models (with both DeePMD and MACE) succeed in preserving the intramolecular geometries and liquid phase behaviour for over 1ns. However, this stability likely indicates that kinetic barriers to break bonds remain high rather than that the MLFFs in question are actually describing the equilibrium phase space of the molecules correctly - that is, we would expect all of these models to fail given sufficiently long simulations. The PC liquid shows a clearer difference between the DeePMD and MACE models: the latter remain stable for long periods, while most of the former fail through unphysical reactions within 100ps. These results indicate that the MACE descriptors recognise the chemical similarity of EC and PC, where the DeePMD descriptors do not. 

Fig.~\ref{fig:OtherMolecules}c shows the change in $t_{\rm stab}$ of each DeePMD model before and after 8 iterations of active-learning - these correspond to the initial and final iterations shown in sec.~\ref{sec:ALresults}. In line with their EC:EMC performance, most models have $t_{\rm stab}\ll 1\,$ns before active learning. Transferring additional training data from GAP improves stability in the same way as for EC:EMC -OPLS+VS performs significantly better than OPLS or OPLS+SM, with the Full training set being slightly more stable still. (In fact, this particular Full model is highly stable across all molecules considered, but that result is unlikely to be general). After active learning the stabilities of most models are significantly improved, suggesting that configurations which correct deficiencies for the EC:EMC liquid are also helpful to correct errors in the other molecules (even VC with its additional functional groups). This result indicates that similar pathological configurations cause trajectory failure in both liquids - we previously argued that these pathologies mostly arise from unphysical close-appraoch of distinct molecules. Including volume scans and running targeted active learning to suppress these high-density local environments is therefore a useful strategy for improving generalisability of an MLIP, as well as performance on its native chemical system.

\section{Conclusions}

This paper has investigated the ability of MLIPs using traditional and graph neural network architectures to describe organic molecular liquids representing conventional battery electrolyte solvents, using very limited training sets and data transferred from a different MLIP architecture. We have argued for the importance of validating models using their native dynamics rather than a fixed validation set, and for comparing collective properties of the system (density, diffusivity) rather than per-atom properties such as energies and forces. We have demonstrated that these collective properties can be very challenging to reproduce, since they are sensitive to the model's description of both the low-energy and barrier regions of the potential energy surface.

Our most important conclusion is that MACE models perform extremely well by all the metrics considered, even for small training sets that contain only classical configurations of the liquid. These models were extremely stable against failures in either intramolecular or intermolecular structure, giving densities that agree well with the reference GAP model. The diffusivities predicted by MACE models were quite different to the reference GAP model, but the tests that we have been able to perform indicate that MACE is probably closer to the correct DFT diffusivities. Unfortunately, the absence of long DFT-MD trajectories makes this assertion difficult to prove conclusively.

By contrast, DeePMD models based on low body-order, short-ranged neural network architectures require larger training sets to achieve good descriptions of a liquid. Networks trained on classical data only dissociate unphysically in very short time intervals. We find that this dissociation results from prediction errors in the intermolecular forces, rather than the intramolecular, and show that it can be largely corrected by training on volume scan configurations that are very cheap to generate. By contrast, transferring active-learned configurations from a different trained model (in our case GAP) provides only minor improvements to model stability and properties, suggesting that the ``holes'' or high-error regions of undertrained configuration space are quite different for DeePMD and GAP.

All the DeePMD models analysed in this work required several iterations of active learning to reach high MD stability, and most achieved good prediction of liquid properties also. Beginning with more transferred data significantly reduced the number of iterations required to reach this performance, allowing us to move quickly into the regime where active-learned configurations improve coverage of the equilibrium configuration space rather than simply correcting predictions of high-energy configurations to prevent model failure. Again, however, transferring active-learned configurations provided little or no improvement in model performance, and we saw indications that more diverse training sets could hinder the ability of the MLIP to capture any one liquid composition accurately. Further work is required to understand how this effect depends on network architecture and training hyperparameters.

Our DeePMD results suggest that data re-use between different MLIP algorithms may not save much computational effort. It is difficult to conclude whether MACE is more able to re-use data, as might be expected given the greater range and generalisability of MACE models, since the MACE models performed very well even with classical data alone. There are interesting suggestions that the density predictions of MACE models containing active-learned configurations are closer to the correct value than those without. 

Finally, we consider how models trained to describe one molecular liquid perform in simulations of a different chemical system. This question is important in modelling battery materials, for example, where chemical properties often depend strongly on the solvent identity, and especially relevant to Foundational Models. Unsurprisingly, we find that small structural changes do not significantly impact model performance, but changing the functional groups degrades model stability significantly (both for DeePMD and MACE). Interestingly, model stability for unseen molecules correlates quite strongly with training set size, even improving on introduction of either ``native'' or ``foreign'' active-learned configurations. This result suggests that any training information that helps to prevent molecular geometry failure will improve the stability of unseen molecules. 

Overall, we conclude that conventional neural network and gaussian process MLIPs remain poorly transferable, and that any such model should be viewed as unique to both its underlying algorithm and its training data with little possibility of transferring data between algorithms. The situation for many-body message-passing models with a longer effective range, such as MACE, is more promising, not least because their data requirements are low enough to provide good performance without substantial active learning or a large number of quantum calculations. Our results suggest that fine-tuning a foundational MLIP with a small amount of system-specific data (whether from a classical forcefield or a pre-existing MLIP) might yield substantial improvements in performance.

A corollary to our conclusions is that training data generated from a foundation model might be usefully transferred to a cheaper MLIP approach, providing initial training sets that compromise between the accuracy of an AIMD calculation and the efficiency of classical MD.

\section*{Conflicts of Interest}

GC has equity interest in Symmetric Group LLP that licenses force fields commercially, and also in Ångstr\"om AI.

\section*{Acknowledgements}

This work was performed using computational resources provided by the Cambridge Service for Data Driven Discovery (CSD3).

\section{Supplementary Information}


\subsection{Molecular structures referred to in the text}


\subsection{Validating diffusion coefficients against reference DFT}

Sec.~\ref{sec:gen0results} showed a systematic difference in centre-of-mass diffusivity between MACE models and the GAP/DeePMD models. We wish to understand which MLIP is closer to the correct value, but evaluating diffusion coefficients using AIMD would be prohibitively expensive. Instead, we validate the predicted energies and forces of each model against the reference DFT method, using configurations extracted from canonical trajectories of the models in question. This approach means that we are not comparing equivalent configurations of the two models, but we are comparing the configurations that contribute to the diffusivity calculation in each case.

We compare two MLIPs: a MACE model trained on the Full dataset, and a DeePMD model trained on the Full dataset with one generation of active learning. (This model was found to give predicted density very close to the GAP reference). We sampled configurations at 20ps intervals from a 1ns constant-volume trajectory at a density of 0.92g/cm$^3$, which is close to the equilibrium density for all three models.

\begin{figure}
	\centering
	\begin{subfigure}[t]{0.48\textwidth}	
    	\subcaptionbox{}{\includegraphics[width=\textwidth]{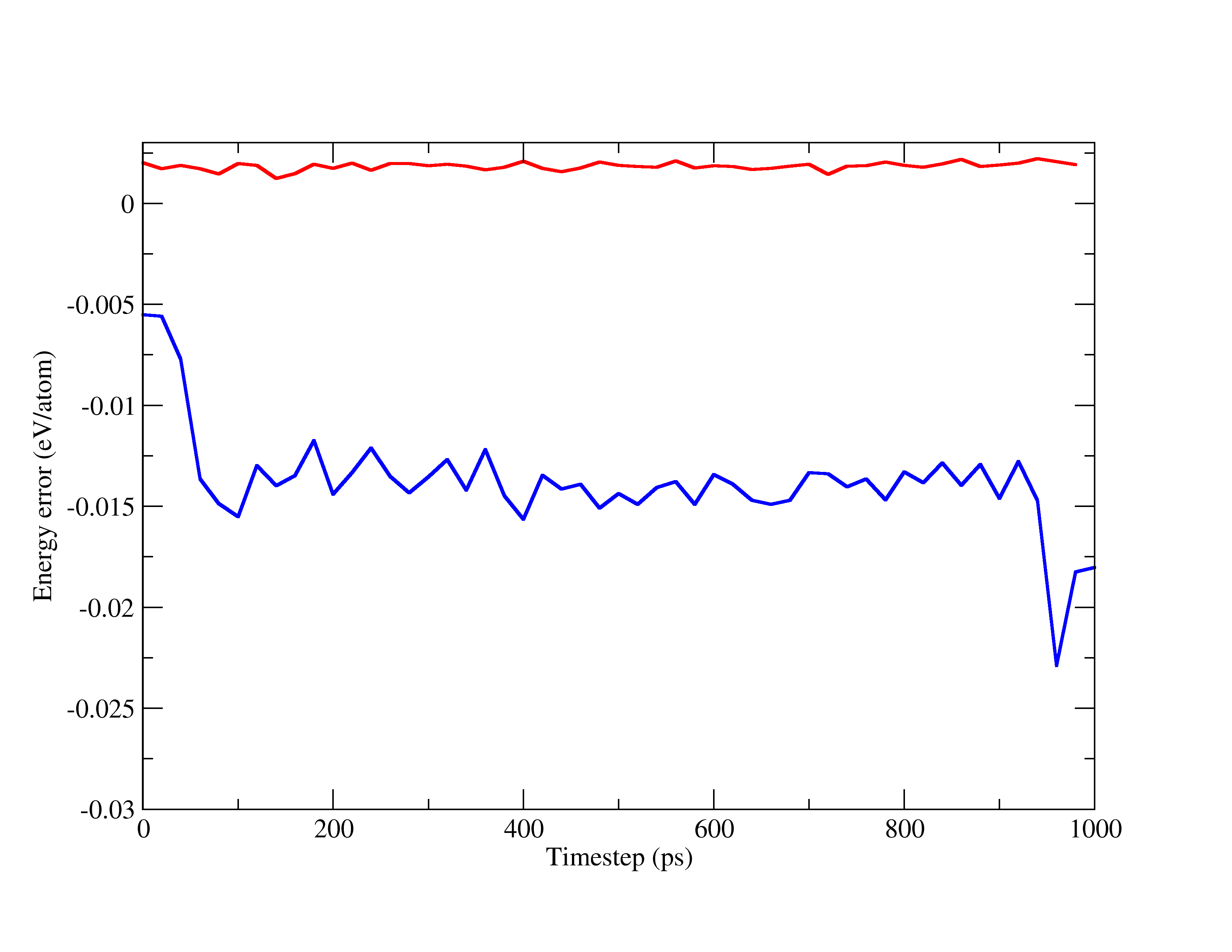}}
    \end{subfigure}	
    \begin{subfigure}[t]{0.48\textwidth}
	\subcaptionbox{}{\includegraphics[width=\textwidth]{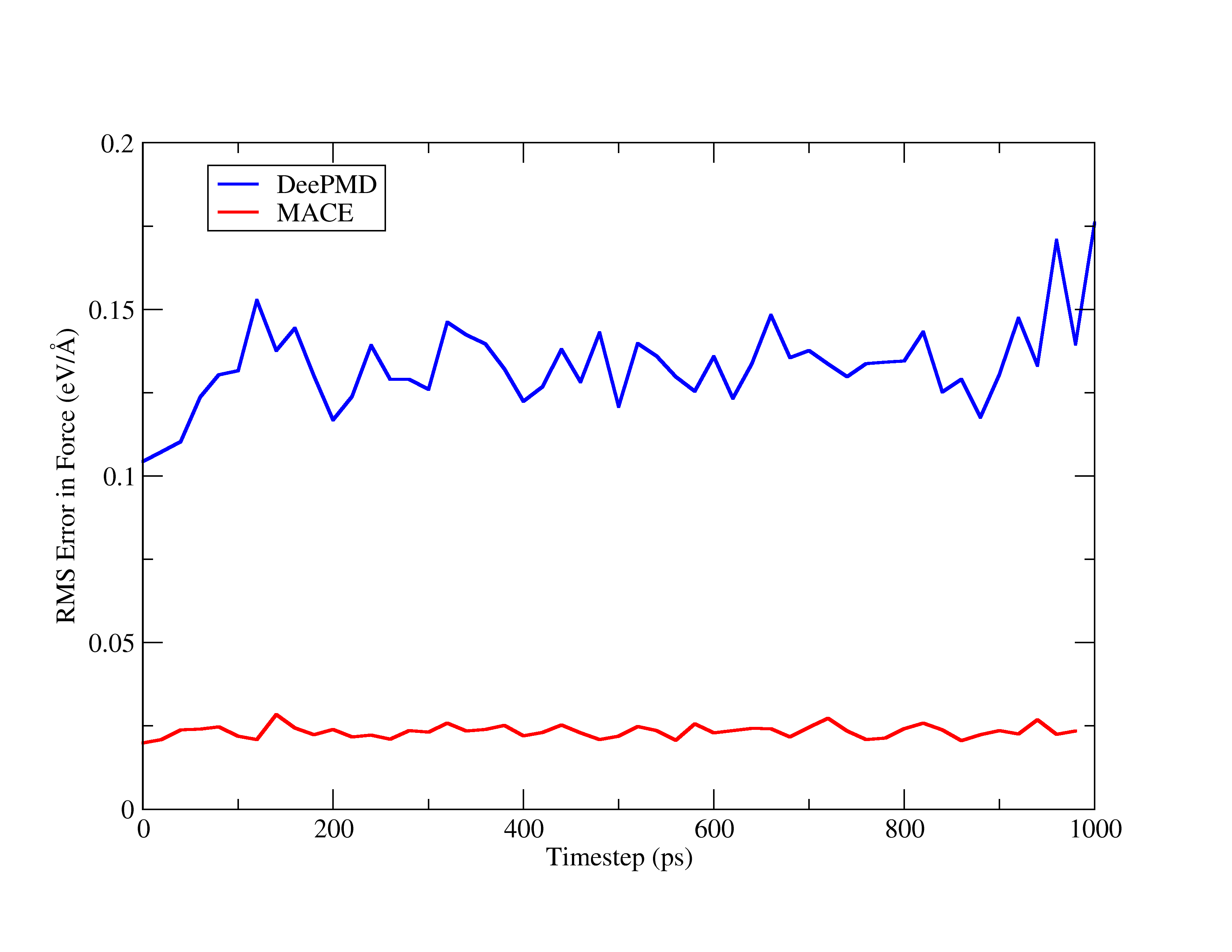}}
	\end{subfigure}	
	\begin{subfigure}[t]{0.48\textwidth}	
    	\subcaptionbox{}{\includegraphics[width=\textwidth]{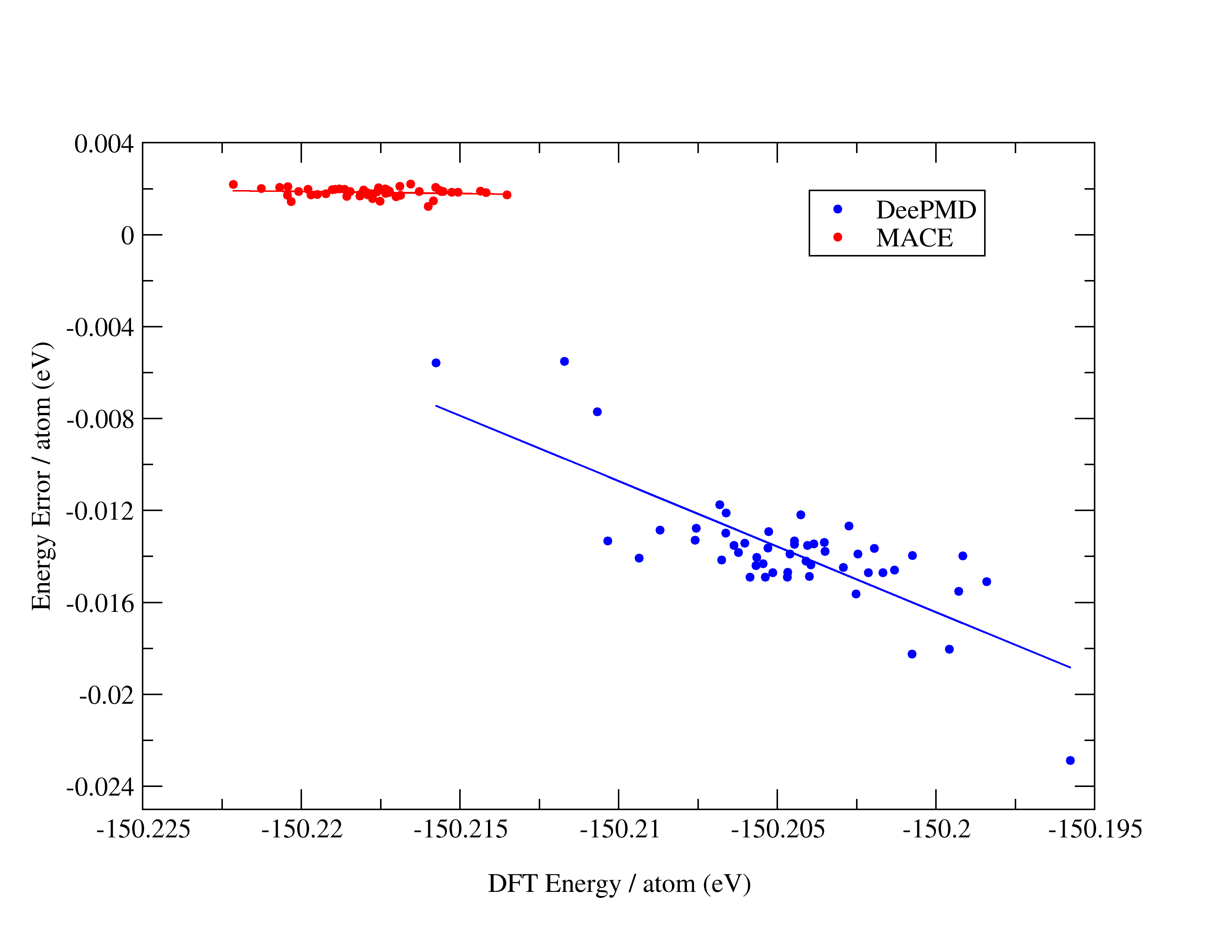}}
    \end{subfigure}	
    \begin{subfigure}[t]{0.48\textwidth}  
	\subcaptionbox{}{\includegraphics[width=\textwidth]{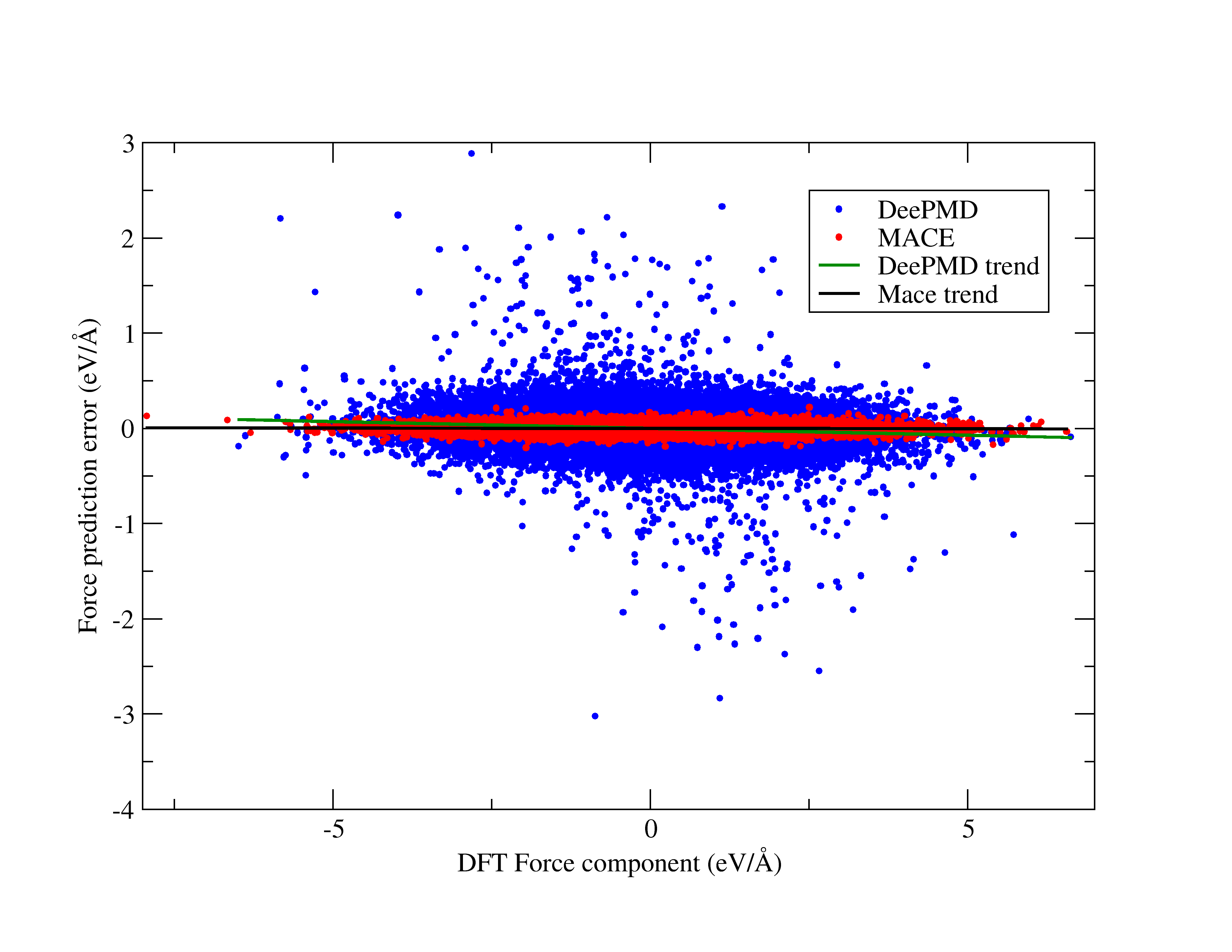}}
	\end{subfigure}
	\caption{Prediction errors for energies (left) and forces (right) in NVT trajectories with comparable DeePMD and MACE models. a) and b) show how the prediction errors change over time during the trajectory, c) and d) show how the errors correlate with the value of the reference DFT value for each quantity.\label{fig:SI_RMSE_NVT}}
   
\end{figure}

Fig.~\ref{fig:SI_RMSE_NVT} shows the prediction errors for total energies and force components. The top panels show that the MACE errors remain small and constant through the trajectory (approx 4meV/atom for the energy and 25meV/\AA for forces, comparable to the equivalent errors for the held-out validation set during training) while the DeePMD errors are larger and increase substantially over the first 100ps. This increase shows that the DeePMD models deviate from their well-trained regions over the course of the simulation, likely decreasing the accuracy of their predicted properties. We therefore expect that the MACE diffusion coefficients will be more accurate.

The lower left panel of fig.~\ref{fig:SI_RMSE_NVT} shows that the DeePMD trajectory is exploring configurations with higher DFT energies than the MACE model, and that these configurations have negative prediction errors (explaining why the model considers them to be easily accessible). Note that this plot does not imply that DeePMD systematically under-predicts the energies of high-energy configurations (since there may be unexplored configurations for which the converse is true) but it does show clearly that the energy barriers being sampled in this trajectory have a much larger energy error than the barriers being sampled by the MACE trajectories.

The lower right panel shows two populations of force components for the DeePMD models: a broad spread of configurations with comparatively small prediction errors, and a smaller population that have much larger errors (comparable with the magnitude of the force component itself). This division agrees with the standard description that configurations far from the training set will experience much higher prediction errors, and also argues that the DeePMD trajectory encounters many such configurations while the MACE model does not.

\newpage

\printbibliography

\end{document}